\documentclass[prd,aps,twocolumn,floats,floatfix,nofootinbib] {revtex4-1}

\usepackage{graphicx}
\usepackage{dcolumn}
\usepackage{bm}
\usepackage{graphics}
\usepackage{slashed}
\usepackage{amssymb}
\usepackage{natbib}
\usepackage{amsmath}
\usepackage{url}
\usepackage[usenames,dvipsnames,svgnames,table]{xcolor}
\usepackage{color}
\usepackage{xcolor}
\usepackage{tabularx}
\usepackage{hyperref}

\newcommand{\trK}{\mathrm{tr}K}
\newcommand{\trT}{\mathrm{tr}T}
\newcommand{\trKhat}{\mathrm{tr}{\hat K}}

\newcommand{\trAtilde}{\mathrm{tr}{\tilde A}}

\begin{document}

\title{Gravitational Wave Signatures of Highly Compact Boson Star Binaries}

\author{
Carlos Palenzuela$^{1}$,
Paolo Pani$^{2}$,
Miguel Bezares$^{1}$,
Vitor Cardoso$^{3,4,5}$,
Luis Lehner$^{4}$,
Steven Liebling$^{6}$,
}

\affiliation{${^1}$Departament  de  F\'{\i}sica $\&$ IAC3,  Universitat  de  les  Illes  Balears  and  Institut  d'Estudis
Espacials  de  Catalunya,  Palma  de  Mallorca,  Baleares  E-07122,  Spain}
\affiliation{$^{2}$ Dipartimento di Fisica, ``Sapienza'' Universit\`a di Roma \& Sezione INFN Roma1, P.A. Moro 5, 00185, Roma, Italy}
\affiliation{${^3}$ CENTRA, Departamento de F\'{\i}sica, Instituto Superior T\'ecnico -- IST, Universidade de Lisboa -- UL,
Avenida Rovisco Pais 1, 1049 Lisboa, Portugal}
\affiliation{${^4}$ Perimeter Institute for Theoretical Physics, 31 Caroline Street North
Waterloo, Ontario N2L 2Y5, Canada}
\affiliation{${^5}$ Theoretical Physics Department, CERN, CH-1211 Geneva 23, Switzerland}
\affiliation{${^6}$Long Island University, Brookville, New York 11548, USA}

\begin{abstract}
Solitonic boson stars are stable objects made of a complex scalar field with a compactness that can reach values comparable to that of neutron stars.
A recent study of the collision of identical boson stars produced only non-rotating boson stars or black holes, suggesting that rotating boson stars may not form from binary mergers.
Here we extend this study to include an analysis of the gravitational waves radiated during the coalescence of such a binary, which is crucial to distinguish these events from other binaries with LIGO and Virgo observations. Our studies reveal that the remnant's gravitational wave signature is mainly governed by its fundamental frequency as it
settles down to a non-rotating boson star, emitting significant gravitational radiation during this post-merger state. We calculate how the waveforms and their post-merger frequencies depend on the compactness of the initial boson stars and estimate analytically the amount of energy radiated after the merger.
\end{abstract}

\maketitle

\section{Introduction}

The direct detection of gravitational waves~(GWs) by LIGO has begun a new era for strong-field gravity. Four events so far \cite{GW150914,GW151226,GW170104,GW170814} appear to represent the inspiral, merger,
and ring-down of binaries composed of two black holes~(BHs), and an additional one is consistent
with a binary neutron star system~\cite{PhysRevLett.119.161101} (or a low mass BH-neutron star 
system~\cite{Yang:2017gfb,PhysRevLett.119.161101}). More detections of compact binary mergers will surely increase our understanding of
compact objects generally.
While BHs and neutron stars now represent the standard model of
compact objects,
exploring the extent to which alternatives differ in their
GW signatures remains an important test.

Among the many exotic alternative compact objects that have been proposed (e.g., fuzzballs~\cite{Mathur:2005zp}, gravastars~\cite{Mazur:2001fv}, wormholes~\cite{Damour:2007ap}, etc),
boson stars~(BSs)~\cite{Ruffini:1969qy} are arguably among the better motivated and very likely
the cleanest to model (see Ref.~\cite{Cardoso:2017njb} for a recent review on exotic compact objects). Indeed, even if perhaps not realized in nature, they
are excellent test-beds to explore possible
phenomenology associated with highly compact objects.

BSs are composed of a complex scalar
field with harmonic time dependence~\cite{Ruffini:1969qy,Colpi:1986ye} (see~\cite{liebpa} for a review). These systems possess a U(1) symmetry that gives rise to a conserved Noether charge. The stability properties of a BS
resemble those of neutron stars, and, in particular, they are stable below a critical mass. The discovery of the Higgs boson in 2012~\cite{Chatrchyan:2012xdj,Aad:2012tfa}
shows that at least one scalar field exists in nature; if other (stable) bosonic particles exist in the universe they might clump together to form self-gravitating objects.

Motivated by the existent and future observations of compact object mergers,
we study here the inspiral and merger of two BSs initially in a tight,
quasi-circular orbit.
The remnant of this merger can generally be
either a BS or a BH. For the merger to result
in a long-lived BS, however, it is required that (i)~the
remnant star be stable, which in turn implies that its mass is less than 
the maximum mass allowed for the model, and (ii)~the angular momentum left over
after the merger satisfy the quantization condition for BSs~\cite{liebpa} and hence either vanish or be larger than the minimum
angular momentum for a rotating BS. 

We are interested in these BS signals within the
context of binary BHs and binary neutron stars, and we will therefore 
consider that range of compactnesses for the initial binary components.
For increasing compactness, the GW signal of a BS binary is expected to
more closely resemble that of a BH binary. For small
compactness, the structure and tidal deformability~\cite{2017arXiv170101116C,Sennett:2017etc} of the
star will play a significant role, similar to the signals from binary
neutron stars. 

This work extends the study begun in~\cite{bezpalen} that 
focused on the dynamics of solitonic BS binaries.
Solitonic BSs~\cite{frie} are a specific family
of BS with a potential (see Eq.~\eqref{potential} below) that yields compactness $C\equiv M/R$ ($M$ and $R$ being the mass and radius of the BS, to be defined below) comparable and even higher than that of  neutron stars.
In particular, the maximum
compactness of the stable configurations can reach values $C_{\rm max}\approx0.33$. As shown in~\cite{car}, a Schwarzschild photon sphere - which is crucial
in order to get a behavior similar to that of a BH under linear
perturbations - appears outside the star at such compactness.

In particular, in~\cite{bezpalen} the constituents of these binaries were solitonic BSs
 with a fixed compactness $C=0.12$ but with different phases and signs in their time oscillations.
For example, the head-on collision of two BSs differing only by a relative phase difference generically produced a massive and significantly perturbed BS. 
However, a BS that collides head-on with another BS with the opposite
frequency but otherwise identical (i.e., an anti-BS) annihilates the Noether charge of the system such that all the scalar field disperses to infinity.

Surprisingly, the addition of angular momentum does not seem to change this qualitative picture~\cite{bezpalen}. Although the merger of a pair of orbiting BSs produces a rotating remnant, the final object eventually settles down into a stationary non-rotating BS by radiating all its angular momentum via GW radiation and scalar field dispersion.
Notice that similar scenarios were studied previously~\cite{pale1,pale2} in the context of mini BSs, which are much less compact than solitonic ones. However, in those cases, the long dynamical timescales of the stars prevented definite conclusions about the end state of the remnants.

Here we focus on the dynamics and GW signatures of the inspiral of binaries consisting of solitonic BSs of increasing compactness. 
As expected,
the frequency of the GWs after the merger
increases with compactness, an aspect they share with
neutron stars and other compact objects. 
We also compare these signals with those of BH and neutron-star binaries and discuss the prospects for differentiating these objects with LIGO/Virgo observations.

This work  is organized as follows. In Sec.~\ref{stp} the structure of the initial data and evolution system are described in some detail. In particular, the Einstein-Klein-Gordon equations are presented, considering a particular potential leading to solitonic BSs. The procedure to construct initial data consisting of an isolated solitonic BS is described, followed by a summary of our initial configurations of binary BS systems. This section culminates with a short summary of the evolution formalism adopted in this work. In Sec.~\ref{cbbs} we study the coalescence of binary BSs for different compactness, focusing  on the dynamics and the corresponding GW signal. Finally, we present our conclusion in Sec.~\ref{con}. Throughout  this paper, Roman letters from the beginning of the alphabet $(a, b,c,\ldots)$ denote space-time indices ranging from $0$ to $3,$ while letters near the middle $(i, j, k,\ldots)$ range from $1$ to $3,$ denoting spatial indices. We also use geometric units in which $G=1$ and $c=1,$ unless otherwise stated.

\section{Setup}\label{stp}

BSs are self-gravitating compact objects made of
a complex scalar field satisfying the Einstein-Klein-Gordon equations (for a review, see~\cite{liebpa}). In this section we present these equations, describe the initial data, detail the evolution formalism, and finally summarize the numerical techniques used to evolve a BS binary.

\subsection{Einstein-Klein-Gordon equations}

BSs are described by the Einstein-Klein-Gordon theory 
\begin{equation}
S=\int d^4 x \sqrt{-g} \left[\frac{R}{16\pi} -g^{ab}\nabla_a\Phi^*\nabla_b\Phi-V\left(|\Phi|^2\right)\right] \,,\label{actionKG}
\end{equation}
where $R$ is the Ricci scalar associated with the metric $g_{ab}$, $\Phi$ is a minimally coupled, complex scalar field, and $V \left(|\Phi|^2\right)$ is the potential for the scalar field. 
Variations with respect to the metric and the scalar field yield the following evolution equations
\begin{eqnarray}
   R_{ab} - \frac{1}{2}g_{ab} R &=& 
   8\pi \, T_{ab} ~~,\\
   g^{ab} \nabla_a \nabla_b \Phi &=& \frac{dV}{d |\Phi|^2} \Phi ~~,
\label{EKG_equations}
\end{eqnarray}
where $T_{ab}$ is the scalar stress-energy
tensor
\begin{equation}
T_{ab} = \nabla_a \Phi \nabla_b \Phi^* +
   \nabla_a \Phi^* \nabla_b \Phi -
   g_{ab} \left[ \nabla^c \Phi \nabla_c \Phi^*
   + V\left(|\Phi|^2\right) \right].  \nonumber
\end{equation}

Different BS models are classified according to their scalar self-potential $V\left(|\Phi|^2\right)$. Here we focus on the so-called solitonic potential~\cite{frie}
\begin{equation}
 V\left(|\Phi|^2\right)=m_b^2|\Phi|^2\left[1-\frac{2|\Phi|^2}{\sigma_0^2}\right]^2\,,\label{potential}
\end{equation}
where $m_b$ and $\sigma_0$ are two free parameters. The bare mass of the scalar field reads as $m_b\hbar$ whereas, in our units, $m_b$ has the dimensions of an inverse mass.
After restoring physical units, the maximum mass of nonspinning BS solutions in this model reads  as $M_{\rm max}\approx 5\left[{10^{-12}}/{\sigma_0}\right]^2\left(\frac{500\,{\rm GeV}}{m_b\hbar}\right)M_{\odot}$, when $\sigma_0\ll1$. Therefore, depending on the model parameters, solitonic BSs can be as massive as (super)massive BHs and, when $\sigma_0\ll1$, they can be slightly more compact than ordinary neutron stars~\cite{frie}.

Owing to the U(1) invariance of the action~\eqref{actionKG}, BSs admit a Noether charge current
\begin{equation}
  J^{a} = i g^{ab} (\Phi^*\,\nabla_{b}\Phi - \Phi\,\nabla_{b} \Phi^*),
\end{equation}
satisfying the conservation law
\begin{equation}
  \nabla_a J^{a} = \frac{1}{\sqrt{-g}} \partial_a 
         \left( \sqrt{-g}\, J^a  \right) = 0 ~.
\end{equation}
The spatial integral of the time component of this current defines the conserved Noether charge, $N$,
which can be interpreted as the number of bosonic particles in the star~\cite{liebpa}.

\subsection{Initial data of binary solitonic BSs}

Our initial data is constructed as a superposition of two boosted, isolated, solitonic BSs.
Let us first describe the solution of a spherically symmetric BS in isolation. The ansatz for the metric and the scalar field are given by
\begin{eqnarray}
\label{staticmetric}
g_{ab}&=&{\rm diag}\left(-e^{\Gamma(r)},e^{\Lambda(r)},r^2,r^2\sin^2{\!\theta}\,\right)\,, \\
\Phi &=& \phi(r)\,e^{-i\omega t}\,,
\end{eqnarray}
where $\omega$ is a real frequency. 
Despite the phase oscillation of the scalar field in time,
the Einstein-Klein-Gordon equations yield solutions that are otherwise static and that are described by the following set of ordinary differential equations
\begin{eqnarray}
{\Lambda}' &=& \frac{1-e^{\Lambda}}{r} + 8\pi r \left(\omega^2 e^{{\Lambda}-\Gamma}{\phi}^2+{\phi}'^2 + e^{\Lambda} V\right),\label{ekg1}\\
\Gamma' &=& \frac{e^{\Lambda}-1}{r} + 8\pi r \left(\omega^2 e^{{\Lambda}-\Gamma}{\phi}^2+{\phi}'^2 - e^{\Lambda} V\right),\label{ekg2}\\
{\phi}'' &=& \left(\frac{{\Lambda}'-\Gamma'}{2}-\frac{2}{r}\right){\phi}'+e^{\Lambda}\left(\frac{d V}{d|\phi^2|}-\omega^2 e^{-\Gamma}\right){\phi} \,, \label{ekg3}
\end{eqnarray}
where a prime denotes derivative with respect to $r$.

The equilibrium configurations are found by integrating numerically
Eqs.~\eqref{ekg1}--\eqref{ekg3} along with suitable boundary conditions. Namely, we impose regularity at the
center, $\Gamma(0)=\Gamma_c$, ${\Lambda}(0)=0$, $\phi(0)=\phi_c$, $\phi'(0)=0$,
and at infinity we require the metric to be Minkowski and the scalar field to vanish.
For a given value $\phi_c$ of the scalar at the center of the star,
the problem is then reduced to an eigenvalue problem for the frequency $\omega$,
which we solve through a standard shooting method. The value of $\Gamma_c$ can be arbitrarily chosen so that $\Gamma(r\to\infty)=0$.
The ADM mass of the BS is $M=m(r\to\infty)$, where $m(r)$ is defined by $e^{-{\Lambda}(r)}\equiv1-{2m(r)}/{r}$.

In the solitonic model, the scalar field has a very steep profile which makes the numerical integration of the equilibrium equations very challenging, requiring very fine-tuned shooting parameters~\cite{mace}.
On the other hand, the steepness of the scalar field also alleviates a customary problem in defining the BS radius; strictly speaking, the scalar field has non-compact support, and thus BSs do not have a hard surface. Following previous work, we define the effective radius $R_M$ as the radius
within which $99\%$ of the total mass is contained, i.e.\ $m(R_M)=0.99M$. In the model under consideration, since the scalar field is very steep, choosing a higher threshold does not affect the mass significantly. Accordingly, we define the compactness as $C\equiv M/R_M$, so that $C=0.5$ for a Schwarzschild BH and $C\approx 0.1-0.2$ for a neutron star. 
In numerical simulations, it is more practical to estimate the radius of the final remnant through the radius that contains $99\%$ of the Noether charge, $R_N$, so we will also adopt this definition when needed. For the initial configurations considered in this work, the two definitions of the radius differ by $5-10\%$, more compact configurations displaying the smaller difference.

\begin{figure*}
\centering
\includegraphics[width=0.32\textwidth]{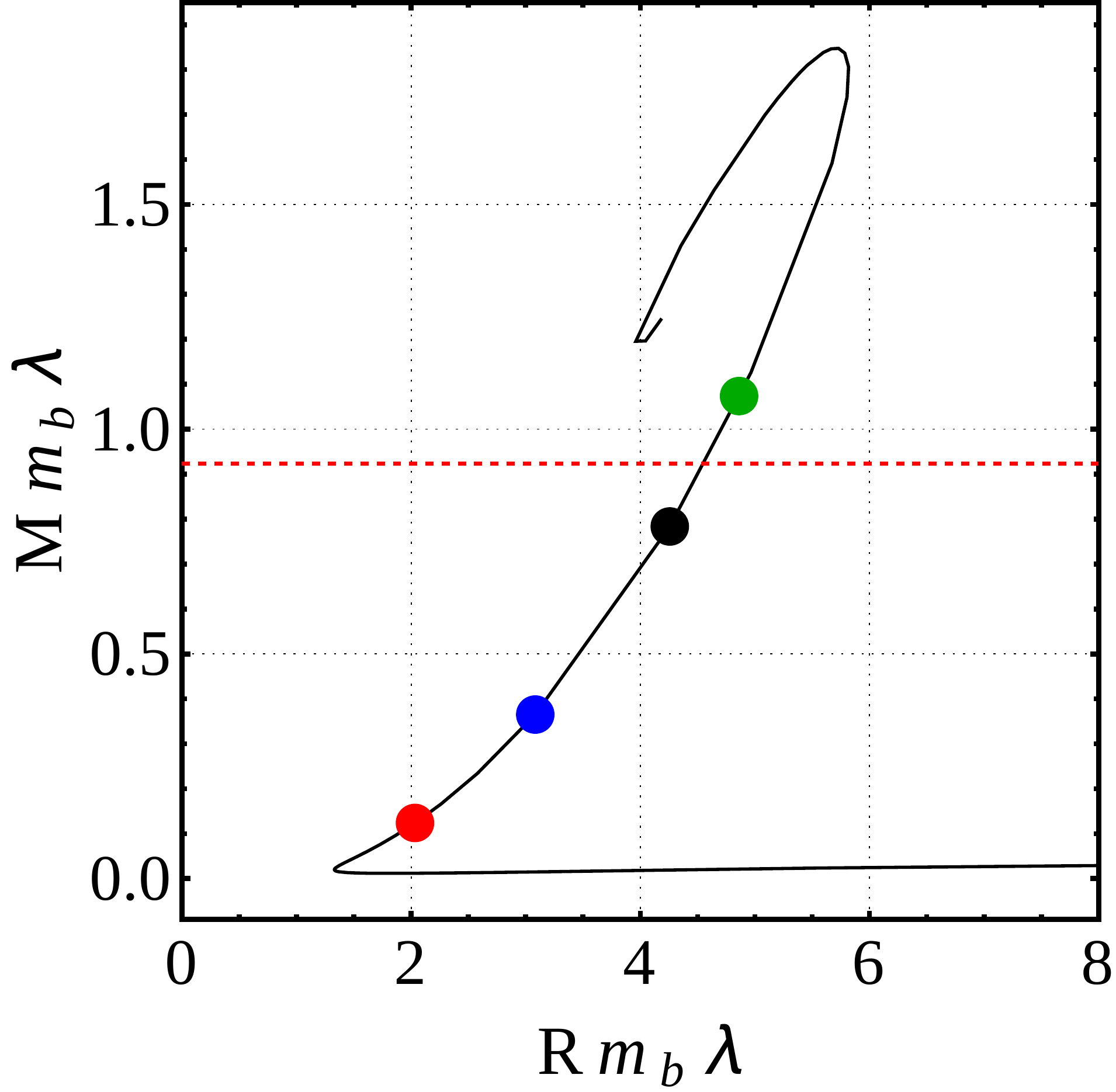}
\includegraphics[width=0.33\textwidth]{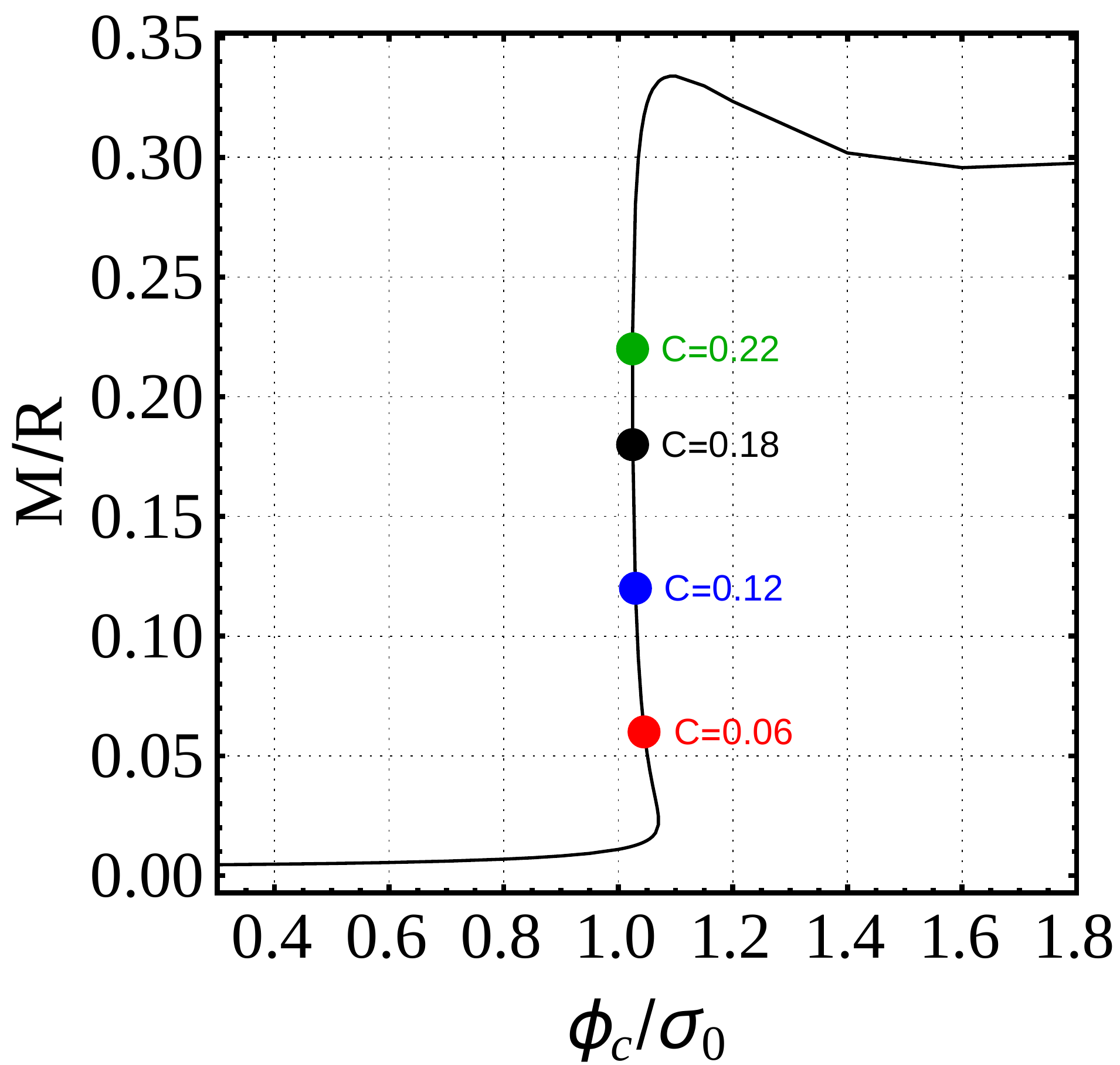}
\includegraphics[width=0.32\textwidth]{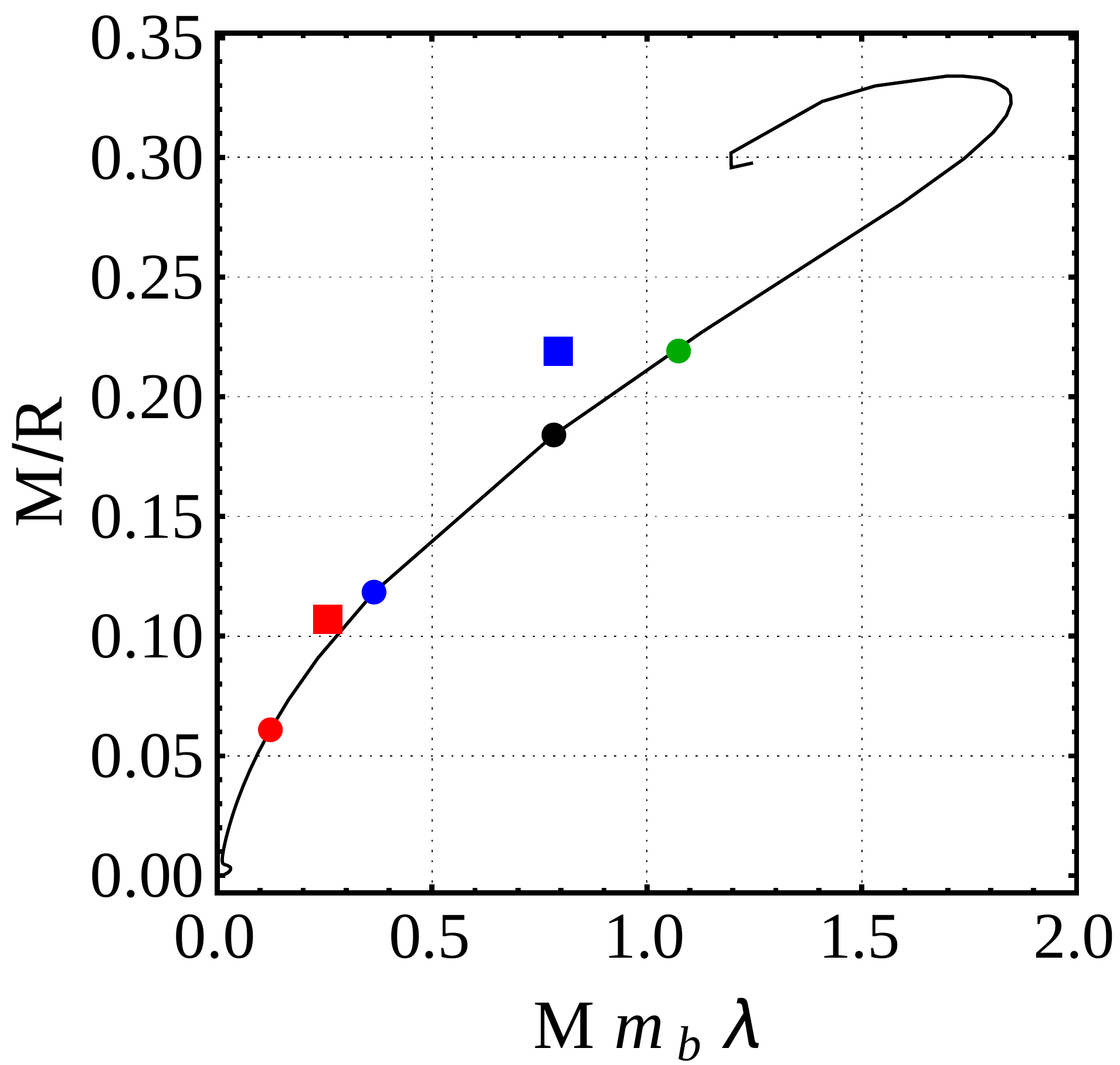}
\caption{{\em Mass-radius diagram (left), compactness as a function of the central value of the scalar field $\phi_c$ (middle), and compactness-mass diagram (right) of an isolated nonspinning BS in the solitonic model~\eqref{potential} with $\sigma_0=0.05$.} Circular markers refer to the \emph{initial} equilibrium configurations considered in this work to construct initial data for BS binaries [cf.\ Table~\ref{isolatedbs}] whereas squared markers in the right panel refer to the final 
remnant produced by the merger of stars in an initial configuration indicated by the same color
[cf.\ Table~\ref{table2}]. 
Two squares corresponding to two configurations are not shown; the remnant of the black configuration is not well-enough resolved and did not reach a quasi-stationary state, and the green configuration produces a BH (with $C>0.5$) instead of a BS.
In the left panel, the horizontal line denotes $M=M_{\rm max}/2$, i.e. the merger of two identical BSs laying above this line is expected to produce a BH (neglecting energy dissipation during the coalescence and non-linear effects).
 The radius $R_M$ is defined as that containing $99\%$ of the mass of the star, except for the radius of the remnant which is instead defined as that containing $99\%$ of the Noether charge, $R_N$.   
}
\label{fig:static}
\end{figure*}

With the above shooting procedure, we compute a sequence of isolated BS solutions
characterized by the central value of the scalar field
$\phi_c$. In particular, here we restrict ourselves to $\sigma_0=0.05$
which is chosen because it allows for very compact, stable configurations.
In Fig.~\ref{fig:static}, a number of relevant quantities of this family
are shown. Among these, we show the compactness $C$ as a function of $\phi_c$ 
which achieves a maximum of roughly $0.33$.
The markers shown in Fig.~\ref{fig:static} denote the four representative initial configurations investigated in this study. The relevant parameters of these solutions are given in Table~\ref{isolatedbs}. 

Following~\cite{mace}, we can rewrite Eqs.~\eqref{ekg1}--\eqref{ekg3} 
in terms of the 
following dimensionless quantities
\begin{eqnarray}
  M (m_b \lambda) ,~~  N (m_b \lambda)^2 ,~~ r (m_b \lambda) ,~~ \omega/(m_b \lambda) \,,
\label{eq:scaling}
\end{eqnarray}
where $\lambda = \sigma_0 \sqrt{8 \pi}$. Doing so produces
equations independent of $m_b$, and hence $m_b$ serves to set
the units of the physical solution. As such, we choose  $m_b$ 
so that the BS mass is $M=0.5$, and the total mass of binary systems constructed
with these solutions (described below) is roughly unity for any compactness of the binary components.
Note that while the scaling with $m_b$ shown in Eq.~\eqref{eq:scaling} 
 is exact,
the scaling with respect to $\lambda$ is not. Only in the $\sigma_0\ll1$ limit
does the scaling hold. For the value $\sigma_0=0.05$ considered here, the  scaling is only approximate.

\begin{table*}
\begin{ruledtabular}
\begin{tabular}{c||ccccc||ccccc|cc}
 $C$ & $\phi_c/\sigma_0$ & $Mm_b \lambda$ & $N(m_b \lambda)^2$ & $(R_M,R_N)m_b  \lambda$ & $\omega/(m_b \lambda)$
   & $m_b$  &  ${M}$ & ${N}$ & $(R_M,R_N)$ & ${\omega}$  & $I/M^3$ & $k_{\rm tidal}$ 
\\ \hline\hline 
 0.06 & 1.045 & 0.1238   & 0.0605   & (2.0334,\,1.8288) & 1.545745909   & 0.9880   & 0.5   & 0.9867  & (8.21,\,7.38)  & 0.3828   & $84.9$ & $8420$\\ 
 0.12 & 1.030 & 0.3650   & 0.2551   & (3.0831,\,2.8360) & 1.066612350   & 2.9124   & 0.5   & 0.4785  & (4.22,\,3.88)  & 0.7787   & $27.8$  & $332$  \\ 
 0.18 & 1.025 & 0.7835   & 0.7193   & (4.2572,\,3.9960) & 0.790449025   & 6.2514   & 0.5   & 0.2929  & (2.71,\,2.54)  & 1.2386   & $12.5$  & $41$ \\ 
 0.22 & 1.025 & 1.0736   & 1.1147   & (4.9647,\,4.7068) & 0.685760351   & 8.5663   & 0.5   & 0.2417  & (2.31,\,2.19)  & 1.4725   & $8.34$  & $20$\\
\end{tabular}
\caption{{\em Characteristics of solitonic BS models with $\sigma_0=0.05$.} The table shows: compactness, central value of the scalar field, ADM mass, Noether charge, radius of the star (i.e, containing $99\%$ of either the mass or of the Noether charge for $R_M$ or $R_N$, respectively) and angular frequency of the phase of $\phi$ in the complex plane, in dimensionless units
on the left and in units such that $M=0.5$ on the right. Note that high-compactness configurations require a very fine tuning in $\omega$. Here we show only the first nine decimal figures. 
In the last two columns, we give the normalized Newtonian moment of inertia (where $I=\int dm L^2$, $L$ being the distance from the axis of rotation) and dimensionless tidal Love number ($k_{\rm tidal}$) of the corresponding configuration as computed in~\cite{2017arXiv170101116C,Sennett:2017etc}.
As a reference, $k_{\rm tidal}\approx 200$ for a neutron star with an ordinary equation of state, and $k_{\rm tidal}=0$ for a BH. 
}\label{isolatedbs}
\end{ruledtabular}
\end{table*}

With the isolated BS configurations in hand, we can now discuss construction of initial data describing an orbiting BS binary. The method, described in detail in~\cite{shi,bezpalen,mundim} is based on a superposition of two boosted isolated BS solutions, and can be summarized as follows:

\begin{enumerate}
\item construct the initial data for two identical, spherically symmetric BSs as discussed in the previous section. We denote the metric and scalar field as $\{ g^{(i)}_{ab}(r), \phi^{(i)}(r)\}$, where the super-index $(i)$ indicates each solitonic star.

\item extend the solution to Cartesian coordinates, with the
center of the star located at a given position $x^j_c$, that is,
     $\{ g^{(i)}_{ab}(x^j - x^j_c), \Phi^{(i)}(x^j - x^j_c,t)\}$.

\item perform a Lorentz transformation to the solution for
each BSs along the $x$-direction with velocity $v_x$, namely
     $\{ g^{(i)}_{ab}{}_\mathrm{boost} \equiv g^{(i)}_{ab}(x^j - x^j_c, v_x), \Phi_\mathrm{boost}^{(i)} \equiv \Phi^{(i)}(x^j - x^j_c, t, v_x)\}$.

\item superpose the two boosted solutions for each of the two stars:
\begin{eqnarray}
\Phi &=& \Phi_\mathrm{boost}^{(1)} + \Phi_\mathrm{boost}^{(2)},\\
g_{ab} &=& g^{(1)}_{ab}{}_\mathrm{boost} + g^{(2)}_{ab}{}_\mathrm{boost} -\eta_{ab}, 
\end{eqnarray}
where $\eta_{ab}$ is the Minkowski metric.
\end{enumerate} 

Notice that this superposition in only an approximate solution that does not satisfy exactly the constraints at the initial time. However,
our evolution scheme enforces the exponential decay of these constraint violation dynamically (e.g., see Figure~10 in~\cite{bezpalen}). 

The positions and velocities of each binary system considered in this work, together with other parameters of our simulations, are presented in Table~\ref{table2}.

\begin{table*}
\begin{ruledtabular}
\begin{tabular}{c||cccc|ccccccc|c}
 $C$ & $y_c^{(i)}$ & $v_x^{(i)}$ & $M_0$ & $J_0$  &
 $t_\mathrm{contact}$ & 
 remnant & $M_r$  & $R^{N}_r$ & $f_r$ & $M_r \omega_r$ & $E_\mathrm{rad}/M_0$ & ${\cal E}_\mathrm{rad}/M_0$
  \\
 \hline\hline 
 0.06 & $\pm 9$ & $\pm 0.140$ & 1.07  & 1.32    &
 950 &
 BS &  0.90 & 8.9 & 0.0207 & 0.117 & 0.075 & 0.029 \\ 
 0.12 & $\pm 5$ & $\pm 0.210$ & 1.18  & 1.24    &	
 300 &
 BS & 0.98 & 4.6 & 0.0311 & 0.203 & 0.085 & 0.057 \\ 
 0.18 & $\pm 5$ & $\pm 0.214$ & 1.29  & 1.40    & 
 330  &
 BS & 1.07 & 2.5 & 0.0489 & 0.329 & 0.120 & 0.086 \\ 
 0.22 &  $\pm 5$ & $\pm 0.220$ & 1.46  & 1.65   & 
 218  &
 BH & 1.42 & -- & 0.0560 & 0.500 & 0.030 & 0.10
\end{tabular}
\caption{{\em Characteristics of binary BS models
and properties of the final remnant}.
The entries of the table are, respectively: the compactness $C$ of the individual BSs in the binary, the initial positions $y_c^{(i)}$, the initial velocities of the boost $v_x^{(i)}$, the initial total ADM mass $M_0$, the initial total orbital angular momentum $J_0$ of the system, the time of contact of the two stars $t_c$, the final remnant, the final total ADM mass $M_r$, the averaged final radius of the remnant star $R^N_r$ (i.e., containing $99\%$ of the total Noether charge), the frequency $f_r$ of the fundamental mode of the remnant, its dimensionless value $M_r \omega_{r}$ (where $\omega_{r}= 2 \pi f_r$), the total radiated energy in gravitational waves for each simulation
$E_\mathrm{rad}$ (i.e., integrated from the beginning and extrapolated to large times after the merger) and the one estimated analytically ${\cal E}_\mathrm{rad}$ as described in appendix~\ref{appA}.
The final angular momentum of the BS remnant tends to zero quite rapidly. The final (dimensionless) angular momentum of the BH obtained in the $C=0.22$ case is $J_r/M_r^2 \approx 0.64$. 
}\label{table2}
\end{ruledtabular}
\end{table*}

\subsection{Evolution formalism}

We adopt the covariant conformal Z4 formulation~(CCZ4) (for full details see~\cite{bezpalen,alic}). The Z4 formulation~\cite{Z44,Z45}  extends the Einstein equations by introducing a new four-vector $Z_a$ which measures the deviation from Einstein's solutions, namely
\begin{eqnarray}
   R_{ab} &+& \nabla_a Z_b + \nabla_a Z_b   = 
   8\pi \, \left( T_{ab} - \frac{1}{2}g_{ab} \,\trT \right) \nonumber \\
   &+& \kappa_{z} \, \left(  n_a Z_b + n_b Z_a - g_{ab} n^c Z_c \right).
\label{Z4cov}
\end{eqnarray}
Notice the presence of damping terms,
proportional to the parameter $\kappa_{z}$, which enforce the dynamical decay of the constraint violations associated with $Z_a$~\cite{gundlach} when $\kappa_{z} >0$. 

To write Eq.~\eqref{Z4cov} as an evolution system, one needs to split the spacetime tensors and equations into their space and time components by means of the $3+1$ decomposition. The line element
can be decomposed as
\begin{equation}
  ds^2 = - \alpha^2 \, dt^2 + \gamma_{ij} \bigl( dx^i + \beta^i dt \bigr) \bigl( dx^j + \beta^j dt \bigr), 
\label{3+1decom}  
\end{equation}
where $\alpha$ is the lapse function, $\beta^{i}$ is the shift vector, and $\gamma_{ij}$ is the induced metric on each spatial foliation, denoted by $\Sigma_{t}$. In this foliation we can define the normal to the hypersurfaces $\Sigma_{t}$ as $n_{a}=(-\alpha,0)$ and the extrinsic curvature $K_{ij} \equiv  -\frac{1}{2}\mathcal{L}_{n}\gamma_{ij}$,  where $\mathcal{L}_{n}$ is the Lie derivative along  $n^{a}$. 

Via a conformal transformation, one transforms to
a conformal metric $\tilde{\gamma}_{ij}$ with unit determinant. In
terms of a real, positive conformal factor $\chi$, one then obtains  a conformal trace-less extrinsic curvature $\tilde{A}_{ij}$ 
\begin{eqnarray}
\tilde{\gamma}_{ij} &=& \chi\,\gamma_{ij},\\
\tilde{A}_{ij}      &=& \chi\left(K_{ij}-\frac{1}{3}\gamma_{ij} \trK \right),
\end{eqnarray}  
where $\trK=\gamma^{ij}K_{ij}$. These definitions leads to new constraints, called conformal constraints, 
\begin{equation}
\tilde{\gamma} = \det (\tilde{\gamma_{ij}})= 1,~~ \trAtilde  =\tilde{\gamma}^{ij}\tilde{A}_{ij}=0,
\end{equation}
which can also be enforced dynamically by including additional
damping terms to the evolution equations proportional to
$\kappa_c >0$.

For further convenience we can redefine some of the evolved
quantities as
\begin{eqnarray}
  \trKhat &\equiv& \trK - 2\, \Theta, \\
  {\hat \Gamma}^i &\equiv& {\tilde \Gamma}^i + 2 Z^{i}/\chi,
\end{eqnarray}  
where $\Theta \equiv - n_{a} Z^{a}$. The final set of evolution
fields is $\{ \chi, {\tilde \gamma}_{ij}, \trKhat, {\tilde A}_{ij}, {\hat \Gamma}^i, \Theta  \}$, and the equations can be
found in~\cite{bezpalen}. These equations must be supplemented
with gauge conditions for the lapse and shift. We use the 1+log 
slicing condition~\cite{BM} and the Gamma-driver shift condition~\cite{alcub}. 

Our version of the CCZ4 formalism, together with the dynamical gauge conditions, is a strongly hyperbolic system, as demonstrated in~\cite{bezpalen}.

\subsection{Numerical setup and analysis}
\label{sec:num}

We adopt finite difference schemes based on the Method of Lines on a regular Cartesian grid. A fourth order accurate spatial discretization satisfying the summation by parts rule, together with a third order accurate (Runge-Kutta) time integrator, are used to achieve stability of the numerical implementation~\cite{cal,Calabrese:2003yd,ander}.
To ensure  sufficient resolution within the BSs in an efficient manner, we employ adaptive mesh refinement (AMR) via the \textsc{had} computational infrastructure that provides distributed, Berger-Oliger style AMR \cite{had,lieb} with full sub-cycling in time, together with an improved treatment of artificial boundaries \cite{lehner}. 
We adopt a Courant parameter of $\lambda_c \approx 0.25$ such that $\Delta t_{l} = \lambda_c \, \Delta x_{l}$ on each refinement level $l$ to guarantee that the Courant-Friedrichs-Levy condition is satisfied. Previous work with this code~\cite{bezpalen,pale1,pale2} established that the code is convergent and consistent for the evolution of BSs.

Our simulations are performed in a domain $[-264,264]^3$ with a coarse resolution of $\Delta x_{1}=4$ and either 7 or 8 levels of refinement, the last one only covering each star, such that the finest resolution is $\Delta x_{8}=0.03125$.

We analyze several relevant physical quantities 
from our simulations, such as the ADM mass, the angular
momentum, and the Noether charge, computed as described in
~\cite{bezpalen}. To analyze the gravitational radiation emitted during the coalescence, we have considered the Newman-Penrose scalar $\Psi_4$. As is customary, we expand $\Psi_4$ in terms of spin-weighted spherical harmonics (with spin weight $s=-2$)
\begin{equation}
 r \Psi_4 (t,r,\theta,\phi) = \sum_{l,m} \Psi_4^{l,m} (t,r) \, {}^{-2}Y_{l,m} (\theta,\phi).
\label{eq:psi4}
\end{equation}
The components $h^{l,m}(t)$ of the strain (per unit distance) can be computed by integrating twice in time the $\Psi_4$. However, this procedure introduces integration constants which might cloud its purely harmonic behavior. These constants do not appear in the frequency domain,
\begin{equation}\label{strain}
  {\tilde h^{l,m}(f)} = - \frac{{\tilde \Psi}_4^{l,m}(f)}{f^2} ~~,
\end{equation}
where ${\tilde \Psi}_4^{l,m}(f) \equiv {\cal F} [{\Psi_4}^{l,m}(t)]$ is the Fourier transform
of the $l,m$ mode of $\Psi_4(t)$. This relation is valid for the physical frequencies $f \ge f_0$, where $f_0$ is the initial orbital frequency. A way to enforce this condition is by setting $f=f_0$ for $f<f_0$ in Eq.~\eqref{strain}~\cite{2011CQGra..28s5015R}, which acts as a high-pass filter and attenuates signals with frequencies lower than the cutoff frequency $f_0$. The strain in the time domain
can be obtained by performing the inverse Fourier transform
$h^{l,m}(t) \equiv {\cal F}^{-1} [{\tilde h}^{l,m}(f)]$.

The GWs, the ADM mass, and the angular momentum are computed as spherical surface integrals at 
different extraction radii to check the consistency of the results.

\section{Coalescence of binary BSs}\label{cbbs}

In this section we analyze the dynamics of binary BSs as a function of their 
individual compactness. For concreteness, we consider four binary systems composed of BSs on the stable branch with compactness ranging from $C\approx0.06$ to $C\approx0.22$  (i.e., see Table \ref{isolatedbs}). Notice that all the stars have been rescaled such that their individual masses in isolation are $M=0.5$, so that the binary has approximately a total initial mass $M_0\approx 1$. 

\subsection{Dynamics}

\begin{figure*}
\centering
\includegraphics[width=17cm, height=15cm]{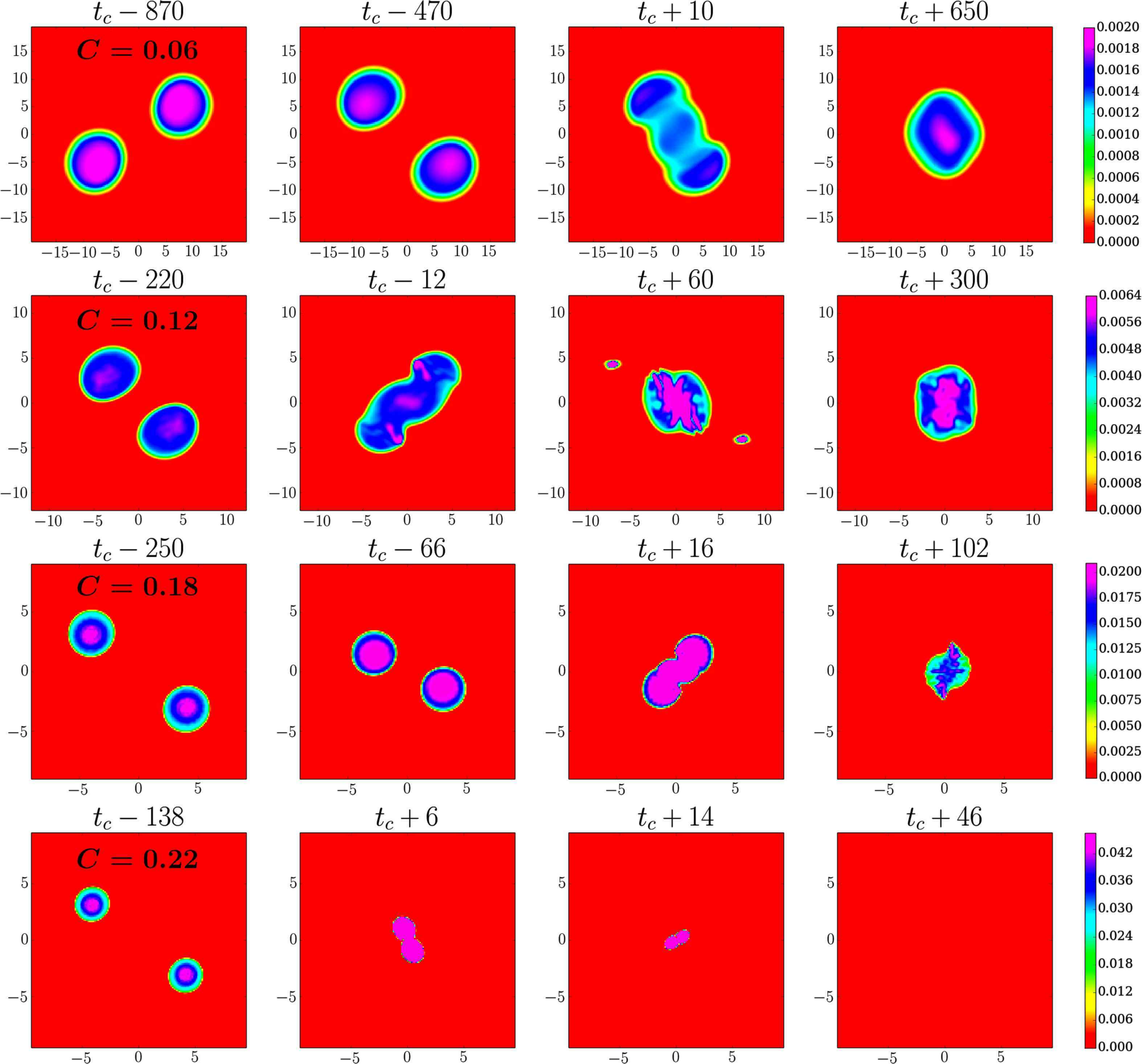}
\caption{{\em Coalescence of binary BSs.} Snapshots in time of the Noether charge density in the orbital plane. 
Each row corresponds to the different compactness (from top
to bottom, $0.06$, $0.12$, $0.18$, and $0.22$). 
The collision of the stars happens at different times due to the different initial conditions and compactness of each case. Note the emission of two scalar blobs in the third panel of the $C=0.12$ case.
}
\label{evol}
\end{figure*}

\begin{figure*}
\centering
\includegraphics[width=17cm, height=15cm]{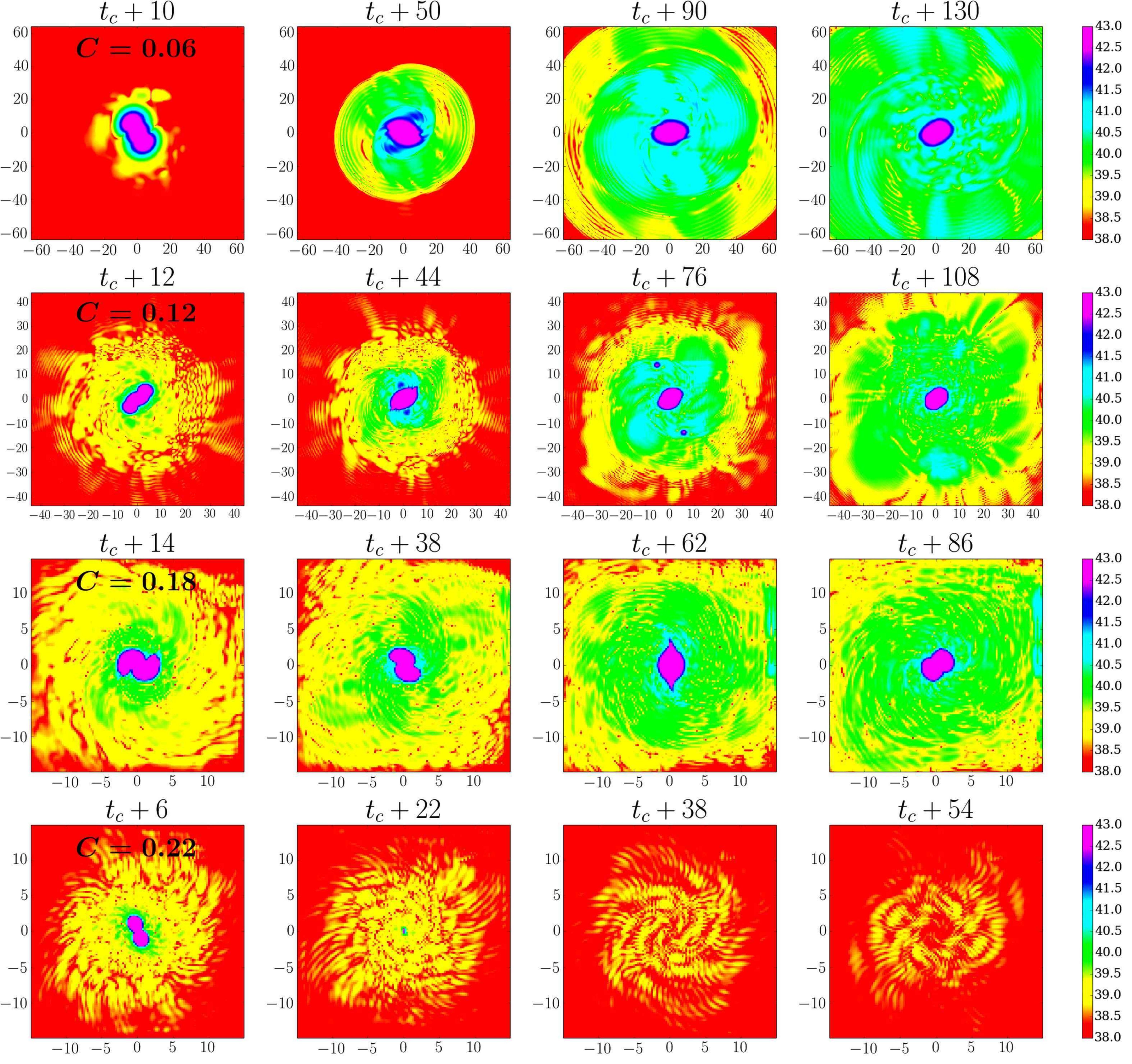}
\caption{{\em Coalescence of binary BSs.} Time snapshots of  $||\phi||$ in the plane $z=0$ in log-scale. Each row corresponds to the different coalescence of BSs cases studied here. In each case the scalar field emitted during the evolution increases after the merger.}
\label{log_evol}
\end{figure*}

Some snapshots of the Noether charge density and the norm of the scalar field during the binary coalescence are displayed in Figs.~\ref{evol} and~\ref{log_evol}, respectively. The early part of the inspiral is qualitatively similar for all cases, with the stars completing at least an orbit before they
make contact. 
Notice, however, that the initial orbital velocities differ among the binaries,
mainly because of two reasons. 
The first reason is that the binaries are constructed
of BSs with different masses, although the scaling is such that the binary mass is approximately unity.
The second reason is that for the smaller compactness cases, the initial 
separation is chosen larger than that for the larger compactness to accommodate the larger size of the stars.
Consequently, the GWs produced during the earlier stages of the inspiral will be weaker than the corresponding stage of the larger compactness cases.

Once the stars make contact, scalar field interactions play a significant
 role in the dynamics of the remnant and nonlinear effects due to the scalar potential becoming significant.
The combination of gravity forces and matter interactions produces a compact, rotating remnant immediately 
after the merger. The final fate of this remnant 
will depend both on the potential and on the location of the solution within the stable branch, that is, on how far from
the unstable branch this specific configuration is. 
This distance can be parametrized in different ways, such as
the fraction $M/M_\mathrm{max}$ (where $M_{\rm max}$ is the maximum allowed mass on the stable branch) or, equivalently,  on the initial stellar compactness. 

Our results indicate a transition between binaries with small compactnesses
with those of larger compactness. In particular, the critical
compactness, $C_T$, appears to be somewhere in the range of $0.18-0.22$.
This transition is roughly estimated by the fact that twice the individual BS
mass above the critical value exceeds the maximum stable BS mass, suggesting
that any remnant (less any radiated scalar field) would be unstable.
We can distinguish two different behaviors depending on the initial compactness, separated by this transition value:

\begin{itemize}
\item $C \lesssim C_T$ : the remnant is a significantly perturbed BS, the angular momentum of which decreases through dispersion of scalar field and 
gravitational radiation, settling down into a non-rotating
BS.
Furthermore, for the case with $C=0.12$, as already observed in Ref.~\cite{bezpalen}, the
angular momentum of the remnant is further reduced through the ejection of two ``rather cohesive''  
scalar field blobs soon after the merger (see third panel, second row in~Fig.~\ref{evol}). 
\item $C \gtrsim C_T $ : the remnant mass exceeds the maximum mass and promptly collapses to a BH with approximately
the mass and angular momentum of the system at the merger time. There is some scalar field surrounding the BH that carries angular momentum and is being either slowly
dispersed to larger distances or falling into the BH. 
\end{itemize}

\begin{figure}
\centering
\includegraphics[width=1.0\linewidth]{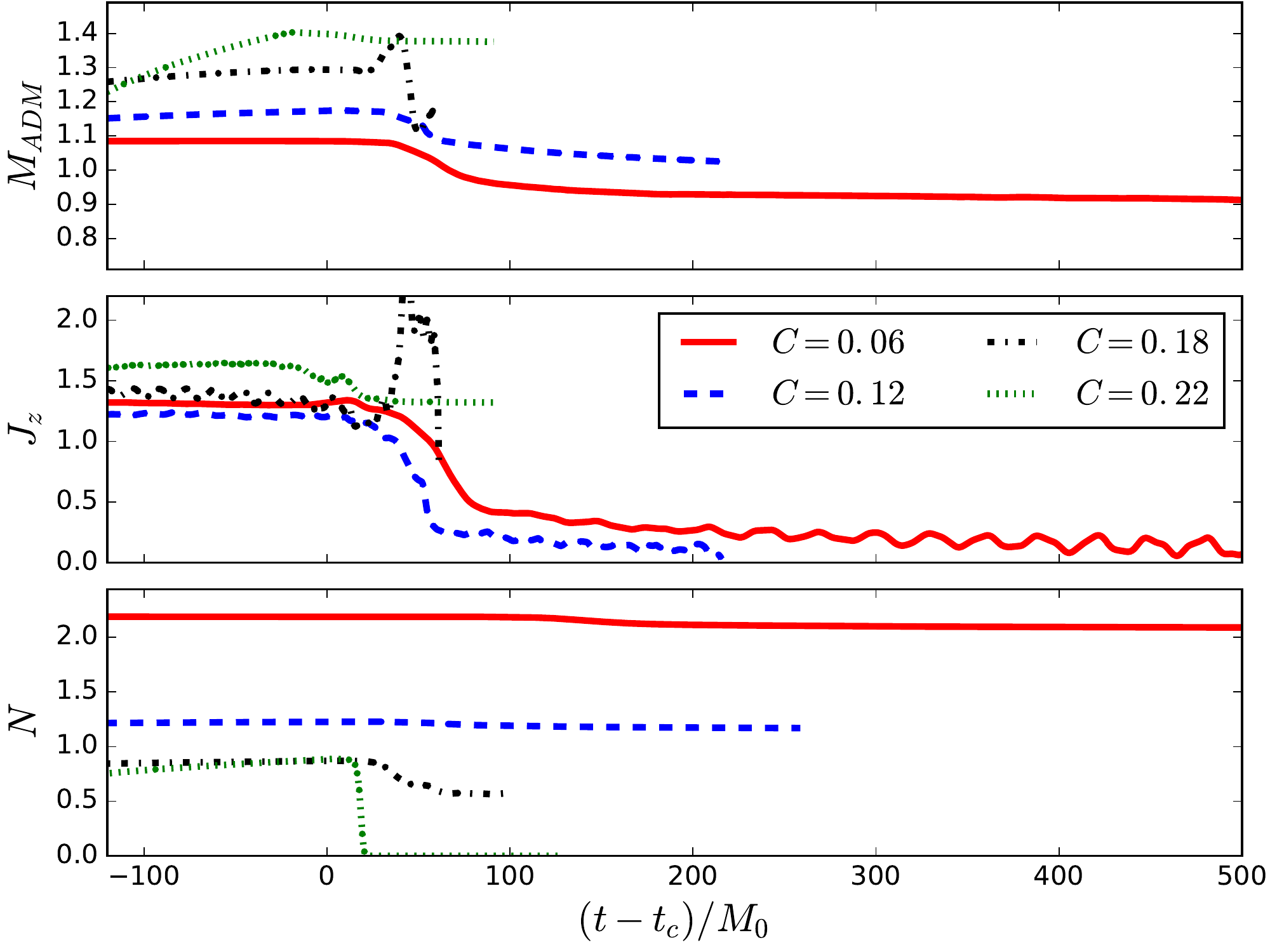}
\caption{ADM mass (top), angular momentum $J_{z}$ (middle) and Noether charge (bottom) as a function on time for the different binaries. During the coalescence the less compact cases (i.e., $C=0.06$ and $C=0.12$) lose only a small percentage  of their initial mass and Noether charge, but almost all their angular momentum. The case $C=0.18$ as
discussed is suspect. The most compact case $C=0.22$ case collapses to a BH after the merger, so the mass and angular momentum do not change
significantly.
} 
\label{admmass}
\end{figure}

The evolution of the ADM mass, angular momentum, and Noether charge are illustrated in Fig.~\ref{admmass}. The binaries show only a significant loss of ADM mass near the merger due to scalar field dispersion/ejection and energy carried away by GWs. Similar behavior is reflected in the total Noether charge --~when the remnant does not collapse to a BH~--. Since the Noether charge is a conserved quantity, the fact that it remains mostly constant further supports the reliability of our  simulations (see also Section~\ref{sec:num} for further discussion of numerical tests).

Notice however that the case $C=0.18$ shows 
peculiar behavior in its mass and angular momentum after
merger when compared to the other cases. A close inspection of the dynamics of this case shows that  large gradients develop that  are not accurately
tracked by the finest resolution we have allowed our adaptive grid to achieve. Indeed, it appears
the system explores a near-threshold regime which is not correctly captured by our simulations and we consider the post-merger period of this case to be
unreliable. Nevertheless, for reference and comparison purposes we include it in the overall discussions since,
in any case, its pre-merger behavior is informative.

Angular momentum during the early inspiral is radiated mainly through GWs. Near the merger stage there is dispersion (and in some cases, also ejection) of scalar field, which also carries away a significant fraction of the angular momentum (notice the sharp decrease in the middle panel of Fig.~\ref{admmass}). After the merger, the final object is much more compact than the initial stars and rotates rapidly, emitting GWs more copiously than the late inspiral. When the remnant is not a BH, the system radiates angular momentum until  settling down to a non-rotating BS. 

As discussed in~\cite{bezpalen}, that the system approaches a non-rotating BS
 might be a consequence of two combined effects: (i)~the quantization of the angular momentum $J_z= k N$ (where $k$ an integer) of rotating BSs that prevents stationary solutions with an arbitrary angular momentum, and (ii)~ 
the rigid structure of the rotation boson star may present difficulties for
the scalar field to organize itself into a stable, rotating configuration.
In particular, the rotating boson star is harmonic both in time and azimuth
with  the level sets of its magnitude being toroidal. This structure
contrasts with a rotating neutron star which can have a range of
rotational profiles either rigid or differential.

Also, another possibility is that the first rotating configuration $(k=1)$ is unstable in the high-compactness regime explored here 
(see Ref.~\cite{Kleihaus:2011sx} for a discussion of the stability of less compact rotating BSs). As shown in Fig.~\ref{fig:static}, the mass and the compactness are very steep functions of the central scalar field, and it is therefore possible that spinning solutions exceed the maximum mass when their non-spinning counterparts are close to such a maximum. If this is the case, the unstable spinning solution would not be formed dynamically and we expect the outcome to be a non-spinning BS or a BH, in agreement with the results of our simulations.

Of particular interest is the case $C=0.12$ due to
the presence of two blobs of
scalar field that are ejected from the remnant at 
speed $v \approx 0.6 c$. The formation of two peaks
in the Noether charge density during the merger is
common for all studied cases, but only for this compactness are they able to detach from
the star while maintaining --~at least temporarily~--
their character (i.e., its Noether charge).
As can be surmised from Fig.~\ref{admmass},
these blobs carry away little mass ($M_{\rm blob}
= m_b N_{\rm blob} \approx 0.025$) but a significant
fraction of angular momentum. A Newtonian estimate,
assuming the distance from the blobs to the plane
of symmetry is $L\approx 7.5$, yields $J_z \approx 2 M_{\rm blob} v L 
\approx 0.2$, consistent with the additional decrease
of angular momentum displayed in the $C=0.12$ case with respect to the $C=0.06$ configuration, as shown in the middle panel of Fig.~\ref{admmass}.

We can gain some insight into the ejection of  these blobs by examining some characteristic speeds
in the problem.
A simple calculation shows that the Newtonian angular orbital frequency $\Omega_c$ when the two identical stars first make contact is 
\begin{equation}
    \Omega_c \approx \frac{C^{3/2}}{2\,M_0}\,,
\end{equation}
where $C$ is the compactness of the individual stars and $M_0$ is the total initial mass of the system. Notice that the blob velocity ($0.6 c$) is considerably larger than the  maximum velocity $v_c= \Omega_c (2 R) \approx 0.35 c$ 
predicted by this rotational rate for solid body rotation.
Similarly one can compute the  velocity associated
with rotation at the angular frequency of the remnant\footnote{remember that there is a factor two between the orbital frequency $\Omega_c$ and the  gravitational one $\omega_c$, namely $\omega_c = 2 \Omega_c$}.
By using the values of table II for the case $C=0.12$ one can obtain that the orbital frequency of the remnant is $\Omega_r = \pi f_r = 0.098$ and its radius is $R_r=4.6$. The velocity of the remnant, $v_r= \Omega_r R_r \approx 0.45 c$,  
also does not reach the level of the blob velocity.

However, these blobs are ejected during the time when the binary is
transitioning from first contact to quasinormal ringing. As such,
the characteristic angular frequency and radius change
$\Omega_c \rightarrow \Omega_r$ and $2 R \rightarrow R_r$. At some point during
 this transition the frequency and the radius might be large enough so that some of the scalar field might move
 with a speed larger than the escape velocity $v_\mathrm{esc} = \sqrt{2 C}$ of the star.
If this occurs, then it is conceivable that  some amount of scalar field may be ejected from the remnant at such a speed.

\subsection{GW signal}

\begin{figure}
\centering
\includegraphics[width=1.0\linewidth]{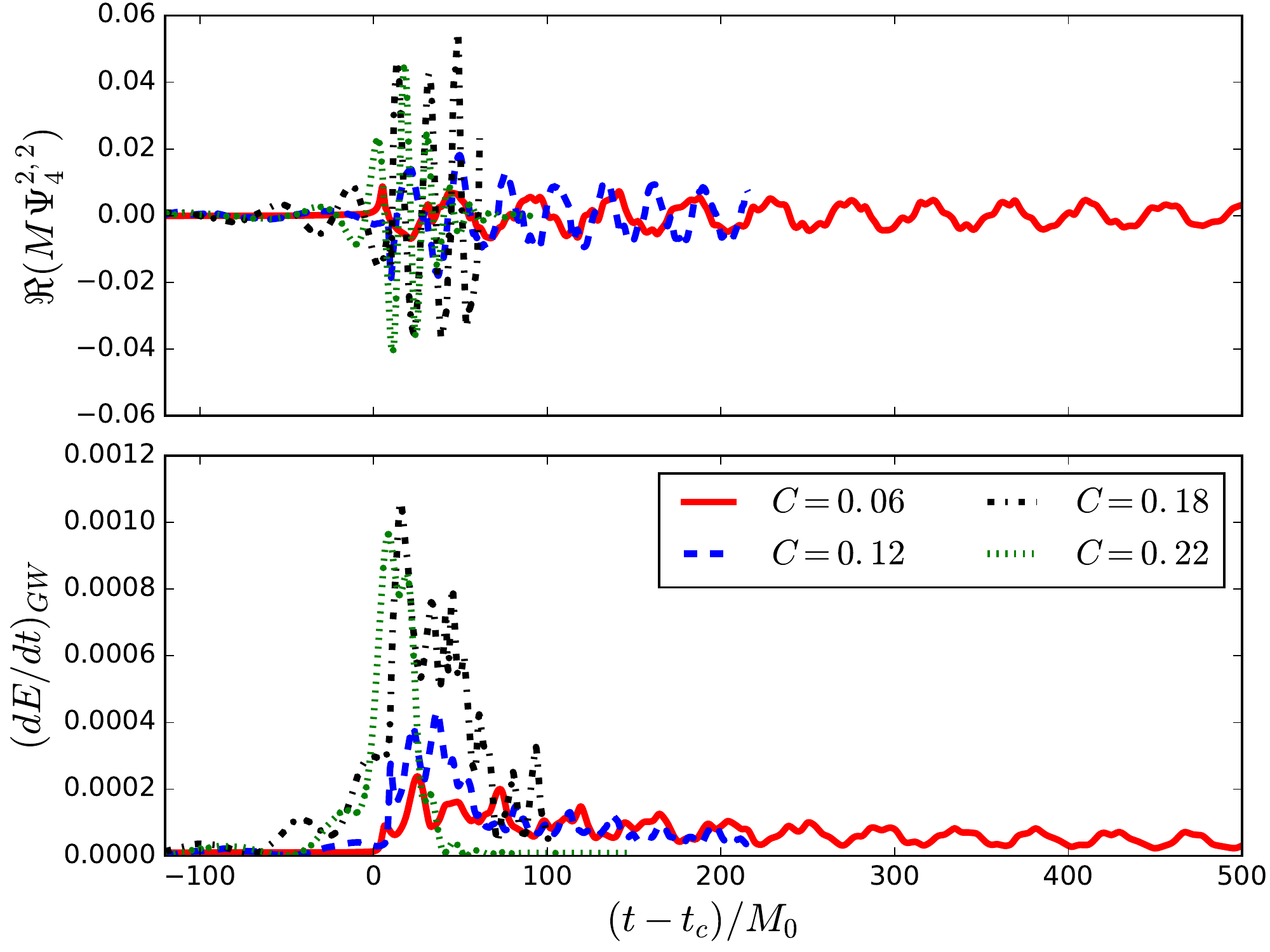}
\caption{ (Top panel) The real part of the main $l=m=2$ mode of the $\Psi_4$ describing
the gravitational emission of the different binaries, as a function
of time.
(Bottom panel) The energy radiated by the main gravitational wave
modes $m=\pm 2$.
} 
\label{psi4}
\end{figure}

\begin{figure}
\centering
\includegraphics[width=1.0\linewidth]{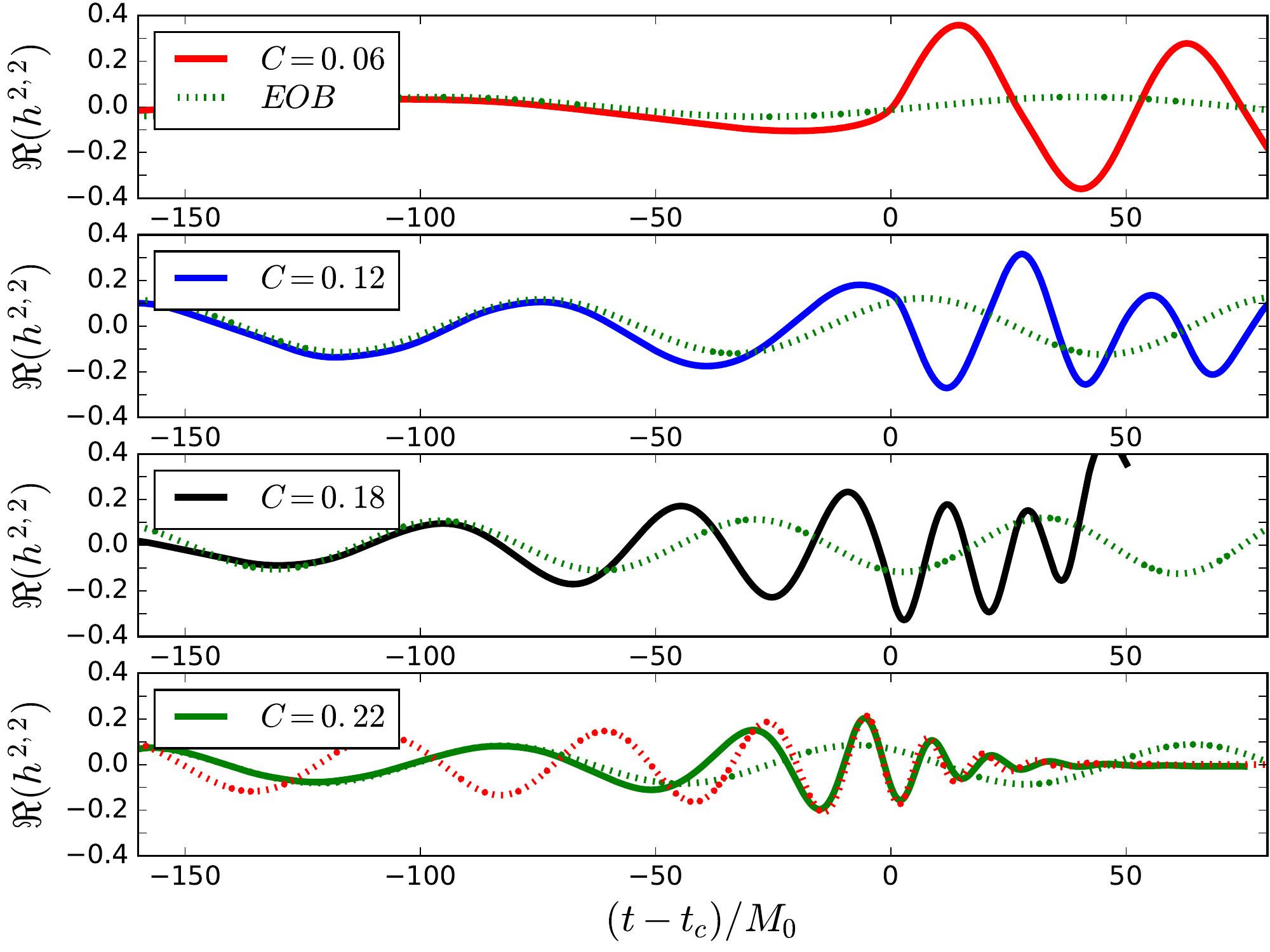}
\caption{The main mode of the strain for 
the different binaries. The time has been rescaled by the initial total ADM mass $M_0$ and shifted such that $t=0$ corresponds to the maximum of the norm of the mode. The amplitude has been also rescaled with the mass of the system. We have chosen the same range in the axes to make clear the increase in frequency as the stellar compactness also increases. The different cases are compared with a recent version of the EOB approximation of a quasi-circular binary BH coalescence~\cite{2017PhRvD..95d4028B} by matching the waveforms at the early inspiral. For the highest compactness $C=0.22$ we have also matched to the EOB waveform at the merger time (dotted red curve).} 
\label{gwstrain}
\end{figure}

The merging binaries produce 
GWs measured by the Newman-Penrose $\Psi_4$ scalar, as displayed in Fig.~\ref{psi4}. In Fig.~\ref{gwstrain} we also show the corresponding strain. Notice that the amplitude and the time scale of the strain has been rescaled with the total initial mass, and the time has been shifted such that the contact time occurs at $t=0$. 

With these waveforms, we can look for the effect of compactness on the gravitational wave signal. Starting with the least compact case ($C=0.06$), it 
radiates the least in the inspiral. The weakness of its inspiral signal results because its stellar constituents have the largest radii and thus they
make contact at the smallest frequency of these four cases. 
As the compactness increases, the late inspiral occurs at higher frequencies and the signal becomes stronger.

Once the stars make contact, both scalar field interactions and gravitational
forces determine the dynamics and tend to homogenize the scalar field profile.
This period is very dynamical producing a rapidly rotating compact remnant that 
radiates strongly in gravitational radiation with an amplitude and frequency
much larger than the inspiral.
This contrast between the pre- and post-merger signals is particularly
marked for the two low compactness binaries, but becomes less so with
increasing compactness. This trend indicates that this contrast likely results from the disparity between the initial compactness of the boson stars and the compactness of the remnant (see Table II).
Notice also that the strain amplitude (Fig.~\ref{gwstrain}) does not show such disparate scales as the Newman-Penrose $\Psi_4$ scalar (Fig.~\ref{psi4}) due to the additional frequency dependence (see Eq.~\ref{strain}).

A simple estimate of the maximum amount of total energy the system can radiate follows from a model of energy balance presented in~\cite{Hanna:2016uhs}, and discussed in detail in Appendix~\ref{appA}.   Within some approximations, when the final object is a non-rotating BS, and the total radiated energy in GWs is estimated to be
\begin{equation}
{\cal E}_{{\rm rad}} \approx 0.96 \, C M \,,
 \label{DeltaErad}
\end{equation}
where $M$ and $C$ are the mass of the system and compactness of the initial stars respectively.
This estimate is largely consistent (i.e., within a factor of two) with the results of our simulations,
given in Table~\ref{table2}, obtained by integrating the gravitational wave luminosity displayed in the bottom panel of Fig.~\ref{psi4}.
Notice the energy emitted in gravitational waves for the case with $C=0.12$ exceeds the $\simeq 5\%$ of the total mass $M_0$ emitted during the analogous coalescence of a binary BH system; thus, boson star binaries in a suitable range can be considered super-emitters in the terminology of~\cite{Hanna:2016uhs}.

The most compact cases considered here are also interesting in the context of the recent observations of GWs by the LIGO detectors. A simple Newtonian calculation shows that the GW frequency at the contact of the two stars is 
$f_c \simeq \Omega_c/ \pi$.
This relation can be contrasted to the frequency at which the analogous case of binary BHs would make a transition
from inspiral to plunge. This frequency is well approximated by the ISCO frequency of the resulting BH produced through the merger~\cite{Buonanno:2007sv}. For non-spinning binary BHs, a handy
expression for the ISCO frequency is provided in~\cite{Hanna:2016uhs}, which indicates that non-rotating BSs would have a contact
frequency higher than the corresponding ISCO frequency for binary BHs provided $C \gtrsim 0.27$ (i.e., a compactness higher than any of the ones considered in this work). 
The gravitational waveforms for the binaries considered here display the expected ``point-particle'' behavior at the 
low frequencies but evidence a sharp transition around contact. 

For cases not collapsing to a BH the gravitational waveforms have a rather
simple structure with principal modes which can be tied to quasi-normal modes of boson stars (see appendices
~\ref{qnm_isolated} and ~\ref{appfinalfeqn}). For cases collapsing to a BH, the post merger gravitational wave signatures  are
captured well by the familiar ring-down behavior of a BH. 
We however note an argument that the angular momentum of the remnant BH is slightly
less than that of the analogous binary BH merger. Because the BH pair
merges well within the system's ISCO, much of the angular momentum is
trapped within the remnant. In contrast, the BS binary, being less compact
and merging at a lower frequency than the BH binary, allows for the radiation
of more angular momentum during the merger.
Recall also that tidal effects introduce modulations in the (late) inspiral waveforms (e.g.~\cite{vines2011post,Sennett:2017etc}) but such modulations become smaller for higher compactness (see the tidal Love numbers in Table~\ref{isolatedbs}). 

\begin{figure}
\centering
\includegraphics[width=1.0\linewidth]{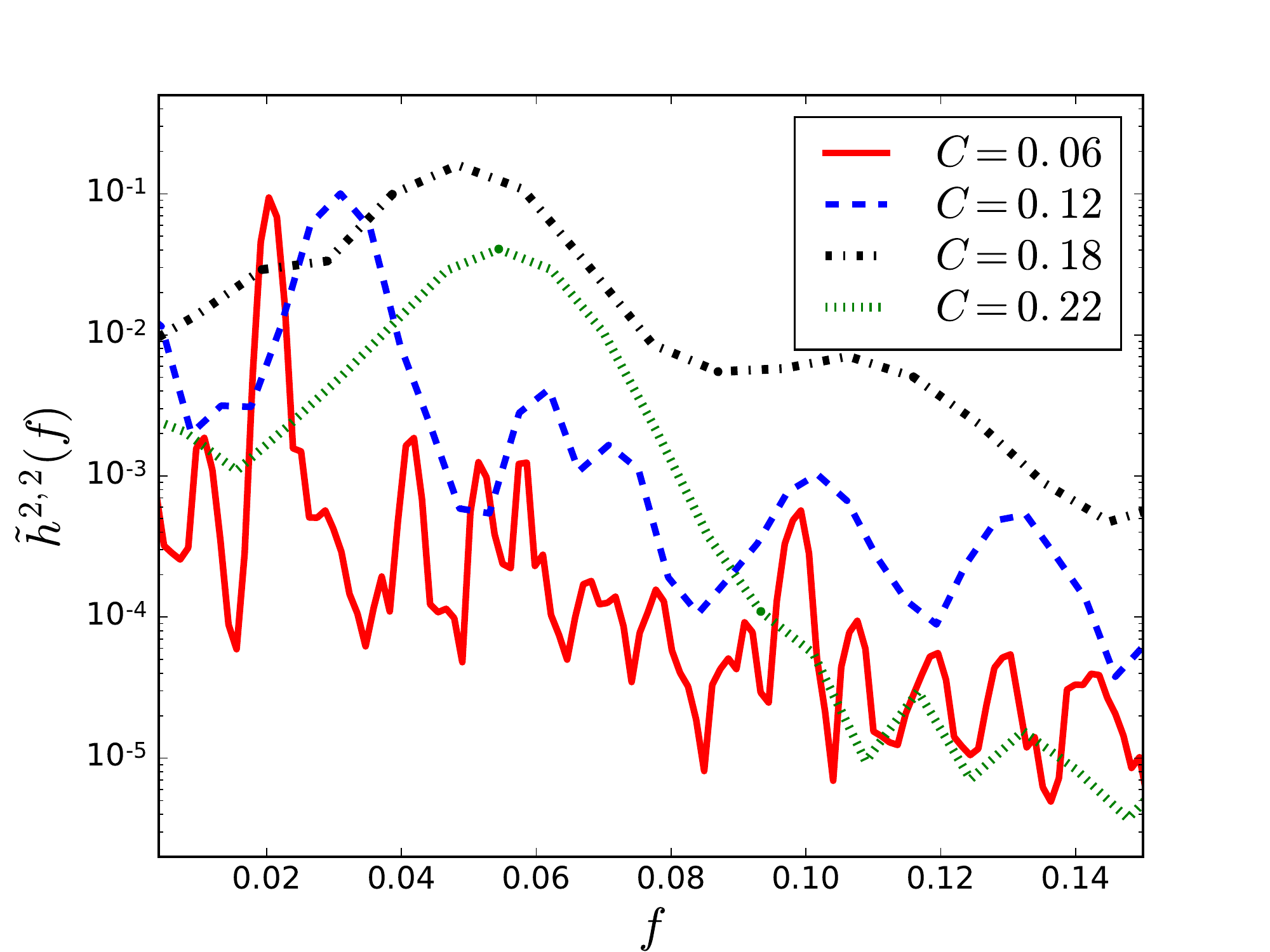}
\caption{Fourier transform of the main mode of the strain
in the post-merger phase, calculated as 
${\tilde h}^{2,2}(f) \equiv {\cal F} [h_{22}(t>t_\mathrm{merger})]$.
Note that peak frequency increases with compactness (similar to neutron stars).
}
\label{Fourier_strain}
\end{figure}

\begin{figure}
\centering
\includegraphics[width=1.0\linewidth]{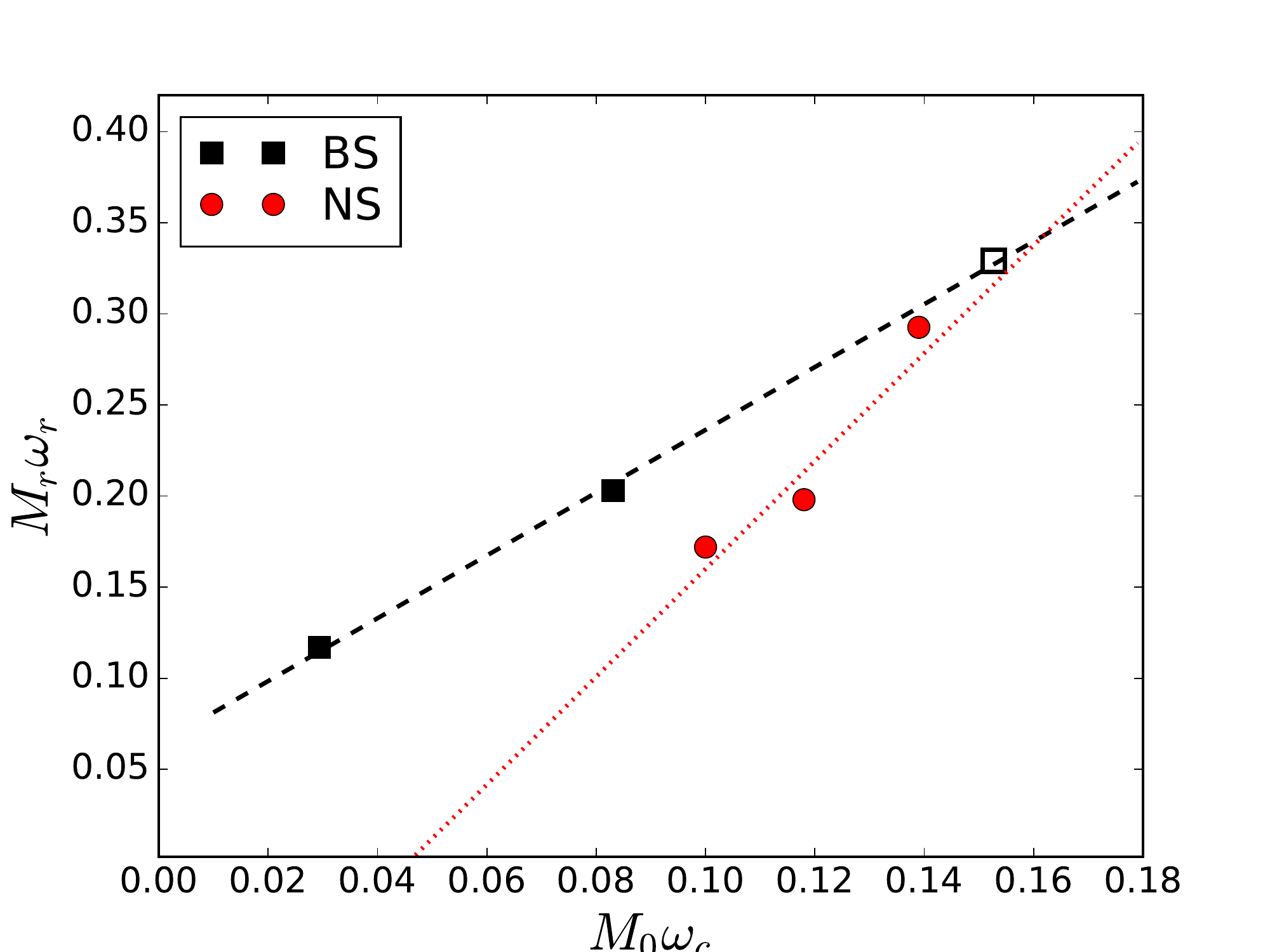}
\caption{ Relationship between the (gravitational) contact angular frequency $\omega_c$ and the (gravitational) angular frequency of the fundamental mode of the remnant $\omega_r$ for the BS binaries considered here. For comparison, we include the neutron star cases studied in~\cite{2015PhRvD..92d4045P}.
The case $C=0.18$ is included
for reference as an unfilled square.
} 
\label{omegas}
\end{figure}

To examine in more detail the after-merger behavior, we analyze the strain in the frequency domain. The Fourier transform of $h_{22}(t>t_\mathrm{merger})$ is shown in Fig.~\ref{Fourier_strain}, where $t_{\rm merger}$ is defined as the time where the strength of the GW is maximum. For $C<C_T$  the final remnant rotates and oscillates while settling down to a (non-rotating) stationary configuration, producing GWs at certain frequencies. Clearly, the frequency of the main peak increases with the compactness of initial objects in the binary. These post-merger frequencies, reported in Table~\ref{table2}, can be fit in terms of the contact frequencies, namely 
\begin{equation}
   M_r \omega_r =  0.064 + 1.72 \, M_0 \omega_{c} \label{bsfit}
\end{equation}
which is valid even when the remnant is not a BH. On the other hand, we can compare
with the fit obtained for neutron stars~\cite{2015PhRvD..92d4045P,2016CQGra..33r4002L}, that in the same units reads
\begin{equation}
   M_r \omega_r =  -0.136 + 2.96 \left(\frac{M}{2.7 M_{\odot}}\right) M_0 \omega_{c} 
\label{nsfit}
\end{equation}
These relations (i.e., with $M=2.7M_{\odot}$ in Eq.~\ref{nsfit}), together with the observed frequencies,  are displayed in~Fig.~\ref{omegas}.
The best fit lines  have quite different slopes and intercepts, but for
high compactness stars they produce similar frequencies. We expect that
the difference in these frequencies implies that remnant
BSs and NSs are potentially distinguishable with GW spectroscopy if 
either (i) a large enough SNR is achieved by increasingly sensitive detectors 
or (ii) a sufficient number of events can be combined (i.e., stacked)~\cite{Yang:2017zxs,Yang:2017xlf}.

Quite interesting is the comparison of the main gravitational wave mode (i.e., $l=m=2$) with the quasi-normal modes of single isolated stars, displayed in Fig.~\ref{qnms_iso_main} and discussed in more detail in Appendix~\ref{qnm_isolated}. Clearly, the frequencies of the remnant agree very well with the frequencies of the fundamental quasi-normal mode of single non-rotating boson stars, providing
further evidence that these cases produce non-rotating, remnant BSs.

This agreement contrasts with the mergers of neutron stars for which the strongest radiating mode is produced by a rotating quadrupole distribution. 
Here, the BS binary remnants are similar to those of binary BHs in that the remnant 
rings down in accordance with QNM dominated by the $m=2$ mode.

\begin{figure}
\centering
\includegraphics[width=1.0\linewidth]{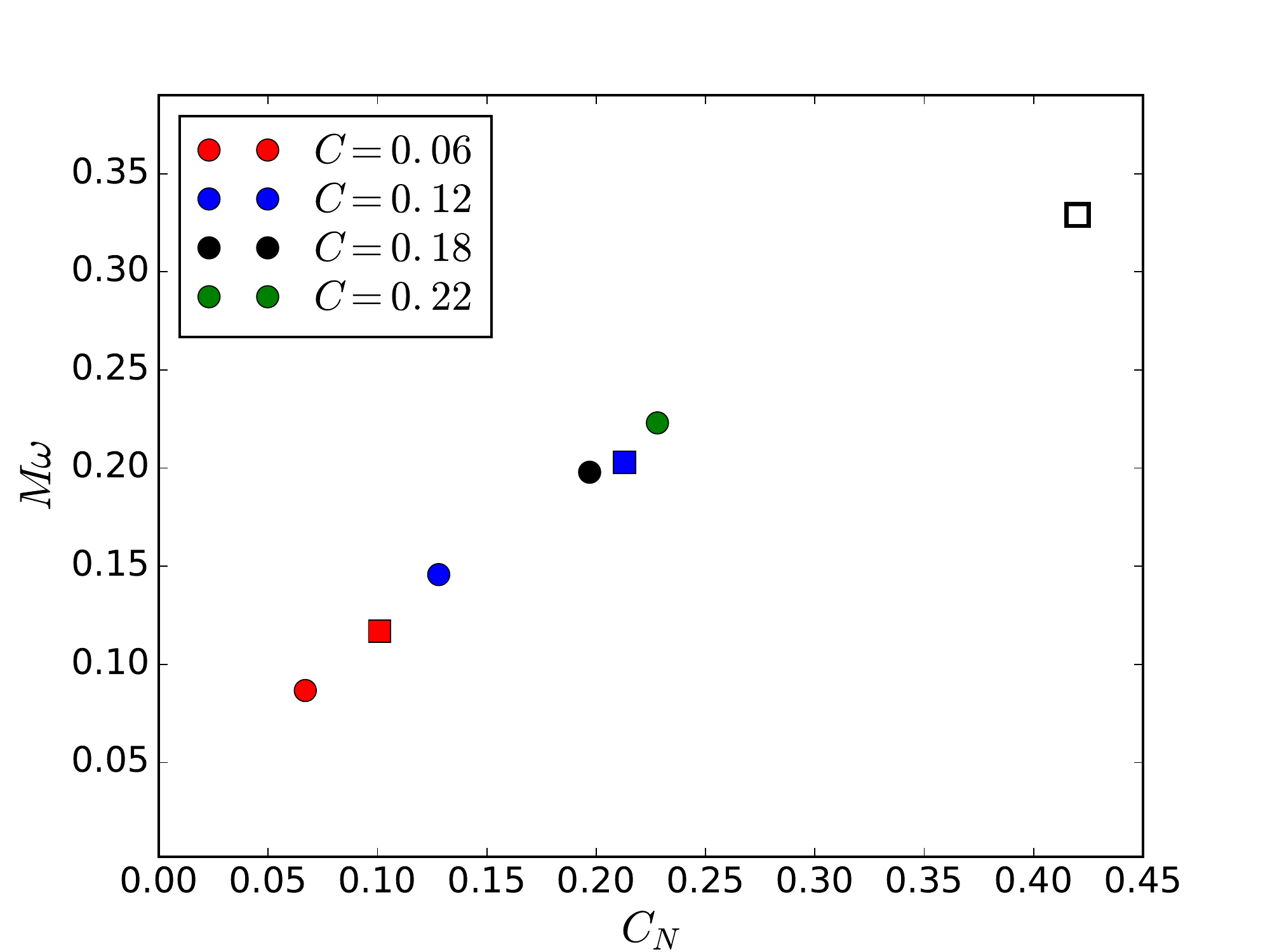}
\caption{ Comparison between the frequencies of the fundamental quasi-normal mode of single BSs in isolation (circles) and the gravitational frequencies of the merger remnant (squares), as a function of the compactness $C \equiv M/R_N$ (the case $C=0.18$ is included
for reference as an unfilled square) where $R_N$ is the radius containing $99\%$ of the Noether charge.
The good agreement between these frequencies suggests that the remnant is indeed a perturbed non-rotating BS ringing down to a quiescent one.
}
\label{qnms_iso_main}
\end{figure}

For $C>C_T$ the final remnant is a {\em rotating} BH, and its post-merger signal shows the characteristic ring-down signal.
For such a case, a significant amount of energy corresponding to the rotational
energy of the BH is retained after merger, which 
contrasts with the cases producing a remnant non-rotating boson star.
For the BH case we can calculate both the frequency and the decay rate of the gravitational wave signature, which can be obtained by fitting the post-merger strain signal to 
\begin{equation}
   h_{22}(t) \approx e^{-\sigma t} \cos(2 \pi f t) ~~.
\end{equation}
We find that $\sigma=0.049 \pm 0.003$ and $f = 0.056 \pm 0.002$.
The final mass and angular momentum of the BH, calculated
asymptotically at a spherical surface of radii $R=50$, are $M_r \approx 1.42$ and $J_r \approx 1.3$ respectively.
Therefore, one can also calculate from linear theory
the quasi-normal-mode frequencies for a BH with final dimensionless spin $a_r = J_r/M_r^2 \approx 0.64$. For this spin, a perturbative calculation yields $M_r \omega_\mathrm{QNM} =0.50819-0.082748 i$~\cite{Berti:2009kk,rdweb}, in very good agreement with the value obtained from the fit $M_r (2 \pi f-i\sigma) \approx 0.5-0.07i$. The small deviations might be due either to inaccuracies in the extraction of the final BH mass or to a sub-leading effect of the scalar field during the post-merger.

Notice that at the end of our simulation there is still some rotating and dispersing scalar field around the BH, but the system does not seem to relax to the hairy BH solution found in~\cite{Herdeiro:2014goa}, at least in the timescales considered here. This is rather natural, if one recalls the {\em quantized} tidal locking relation $\omega=m\Omega_H$, which is necessary for the existence of such solutions (with $m= \pm 1, \pm 2,...$). This condition relates the internal frequency of the BS, $\omega$, with the angular velocity of the BH horizon, $\Omega_H\equiv J_r/(2 M_r^2 R_H)$ (where $R_H$ is the radius of the horizon). In our simulations $\Omega_H \approx 0.087$ and
$\omega \in [0.4-1.8]$, and thus $m$ should be between $4-20$. Since the equal mass binary systems drive primarily the
$m=2$ mode, it is difficult to fulfill the tidal locking condition unless higher modes are 
nonlinearly generated. Though,
these would be significantly sub-leading in strength.
%

\section{Conclusions}\label{con}

In this work we have studied in detail the dynamics of binary BSs for different initial compactness in the range $[0.06,0.22]$. Our analysis reveals several interesting features. We confirm and extend the results obtained 
in~\cite{bezpalen}, namely that the merger of two solitonic BSs leads either to a BH or to a non-rotating BS. The latter
case is due to the inherent difficulty to achieving, simultaneously, the quantized angular momentum condition, the specific profile for the scalar field required in rotating boson stars, and stability of highly-compact rotating BS.

Gravitational waves emitted
during the coalescence of these objects show that for low-compactness BSs ($C < C_T$), the maximum strength achievable
in the inspiral phase is rather weak but it rises rapidly during merger with a significant amount of energy during that phase.
For high-compactness BSs ($C > C_T$), a more monotonic transition --as judged in the rate of upward frequency sweep--
is observed in between inspiral and merger phases. The final object  promptly collapses to a BH and the post-merger
gravitational wave is dominated by the typical ring-down of a spinning BH. 

For the less compact cases with $C<C_T$, the main mode in the post-merger gravitational radiation is given
by the fundamental quasi-normal frequency of the non-rotating boson star. This is in contrast with the behavior manifested
in binary neutron stars (that do not collapse promptly to a BH), where the main mode is linked to the rotation
of the newly formed (hyper) massive neutron star. Nevertheless, in both cases, the main mode can be linked to the
contact frequency by a rather simple linear relation. Importantly for efforts to try and distinguish binary boson stars
from binary neutron stars with gravitational waves, the relation is sufficiently distinct to be probed by 
3rd-generation detectors and/or the combination of multiple events in aLIGO/VIRGO.

Moreover, it is clear that the overall gravitational wave signals from solitonic boson stars 
are unlikely degenerate with those from either binary neutron stars or binary BHs except possibly
in the highest compactness cases (see related discussion in~\cite{Yunes:2016jcc}).
That is, if the primary and secondary objects are not compact enough, the late inspiral phase will differ significantly
from that of two BHs and the remnant will have peak gravitational modes at significantly higher frequencies in
general than those expected in binary neutron star mergers (which do not collapse to a BH promptly). Thus, sensitive searches could in principle distinguish rather easily these two systems. 
For compactnesses of the order of $C_T \gtrapprox 0.2$ a BH is formed but one with less angular
momentum than a corresponding BH binary would have formed. 
As such, with a sufficiently sensitive detection of the post-merger
spin, one may be able to differentiate a compact BS binary from a BH binary
by comparing to the inspiral angular momentum. Notice however that there could still be
a degeneracy with a sufficiently compact binary neutron star merger, 
the presence of an electromagnetic counterpart would favor the merger involving at least
one neutron star as opposed to binary boson stars~\cite{2041-8205-848-2-L12}.

We also notice in passing that in all cases considered in this work (and likely in all cases allowed by the solitonic model) the final BS remnant is not compact enough to produce echoes in the post-merger ring-down phase at late time~\cite{Cardoso:2017cqb}, since this effect requires the exotic compact object to have an external, unstable, photon sphere, which is not the case for the BS models discussed here.

We expect our results to be qualitatively similar in other scenarios including self-gravitating, compact scalar configurations. For instance, if the scalar field is real, action~\eqref{actionKG} admits compact, self-gravitating, oscillating solutions known as oscillatons~\cite{Seidel:1991zh}. These solutions are metastable, but their decay time scale can significant exceed the age of the universe and we expect their properties to be very similar to those of BSs. 

Although not likely to have astrophysical consequences, it would nevertheless be interesting to investigate the transition at $C = C_T$ which separates markedly different behaviors in the post-merger phase. 

\subsection*{Acknowledgments} 

We would like to thank Xisco Jimenez for helpful discussions and for providing the EOB waveform.
CP and MB acknowledge support from the Spanish Ministry of Economy and Competitiveness grants FPA2013-41042-P and AYA2016-80289-P (AEI/FEDER, UE). CP also acknowledges support from the Spanish Ministry of Education and Science through a Ramon y Cajal grant. MB would like to thank CONICYT Becas Chile (Concurso Becas de Doctorado en el Extranjero) for financial support. PP acknowledges financial support provided under the European Union's H2020 ERC, Starting Grant agreement no.~DarkGRA--757480. LL thanks CIFAR and NSERC for support.
VC acknowledges financial support provided under the European Union's H2020 ERC Consolidator Grant ``Matter and strong-field gravity: New frontiers in Einstein's theory'' grant agreement no. MaGRaTh--646597. Research at Perimeter Institute is supported by the Government of Canada through Industry Canada and by the Province of Ontario through the Ministry of Economic Development $\&$
Innovation.
We acknowledge networking support from COST Action CA16104 ``GWverse'', supported by COST (European Cooperation in Science and Technology).
This project has received funding from the European Union's Horizon 2020 research and innovation programme under the Marie Sklodowska-Curie grant agreement No 690904
This work was supported in part by the NSF under grant PHY-1607291~(LIU). We thankfully acknowledge the computer resources at MareNostrum and the technical support provided by Barcelona Supercomputing Center (RES-AECT-2017-2-0003).
Computations were also performed on XSEDE computational resources.

\appendix

\section{Estimate of the total gravitational radiation in the post-merger stage}\label{appA}

A simple estimate of the amount of energy the system can radiate 
can be obtained following the model presented in~\cite{Hanna:2016uhs}. Aided
by the results from numerical simulations, we can further refine such a model for
the behavior observed in the merger of binary BSs. Namely, we observe that, in the case
where collapse to a BH is avoided, the final result is a BS remnant with no angular momentum and with radius $R_r$ having mass $M_r$ that is roughly the total initial mass $M_0$ (i.e., $M_r \approx M_0 = M_1 + M_2$) . The merger takes place at ``contact," that is, when the stars are separated by $R_1+R_2$. Thus, the energy of the system
at such an instant is roughly
\begin{eqnarray}
E_\mathrm{contact} &=& E^\mathrm{pot}_{12} + E^\mathrm{kin}_{12} +
              E^\mathrm{pot}_{1} +  E^\mathrm{pot}_{2} \nonumber \\
           &=& - \frac{M_1 M_2}{R_1 + R_2} + \frac{1}{2} I_c \Omega_c^2
              - \frac{3 M_1^2}{5 R_1}  - \frac{3 M_2^2}{5 R_2}  ~~,~~
\end{eqnarray}
where we have included the binding energy of each star and have assumed that the stars have constant density. The moment of inertia $I_c$ with respect to the center of mass, assuming the stars are irrotational, can be written at contact time as
\begin{equation}\label{Ic}
 I_c \equiv  \frac{M_1 M_2}{M_1 + M_2} (R_1 + R_2)^2       .
\end{equation}
Notice that at contact time the orbital frequency can be
estimated as $\Omega_c^2 = \frac{M_1 + M_2}{(R_1+R_2)^3}$.

Following the discussion
in~\cite{Hanna:2016uhs}, for the equal mass case
($M_1=M_2=M$ and $R_1=R_2=R$ ) this energy can be expressed as a function of the compactness $C=C_1=C_2$ as
\begin{equation}
    E_\mathrm{contact} \approx - \frac{29}{20} M C.
\end{equation}
As the collapse takes place, the system 
ultimately settles into a non-rotating BS. The energy
left in the system (beyond the rest mass) is given by the potential energy of the BS which, assuming a spherical object of uniform density, can be estimated as
\begin{equation}
E_\mathrm{final} = -\frac{3 M_r^2}{5 R_r}  
         = - \frac{12}{5} M C \frac{R}{R_r} ~~,
\end{equation}
where we have considered an upper bound $M_r \approx 2 M$.
Assuming no scalar radiation, we can now estimate the radiated energy in gravitational waves 
during different states of the system. In particular,
the total amount of energy radiated ${\cal E}_{{\rm rad}}$ and radiated after contact ${\cal E}^{ac}_{{\rm rad}}$ are,
\begin{eqnarray} 
{\cal E}_{{\rm rad}} &=& - (E_\mathrm{final} - E_\mathrm{initial}) \\
&\approx&  \frac{6}{5} (2 \frac{R}{R_r} -1 )  \, C M  \\
&\approx&  0.96\, C M = 0.48\, C M_0~~,
\end{eqnarray}
and
\begin{eqnarray}
{\cal E}^{ac}_{{\rm rad}} &=& - (E_\mathrm{final} - E_\mathrm{contact}) \\
 &\approx&
 \frac{M C}{20} \left(-29 + 48 \frac{R}{R_r}  \right) \\
 &\approx&  0.71 \, C M = 0.36 \, C M_0 ~~,
\end{eqnarray}
and we have estimated the ratio $R/R_r\approx 0.9$ from our simulations (alternatively, assuming the effective density of the individual BSs is similar to that of the merged BS, one has  $R/R_r = 1/2^{1/3}\approx 0.8$). Notice
that for $C\gtrapprox 0.1$, if no collapse to a black hole takes place, this implies that highly compact
binary boson star systems are examples of ``super-emitters,'' in the terminology introduced in~\cite{Hanna:2016uhs} 
(as the analogous BH binary system emits $\approx 5\%$ of their total mass).

\section{Estimate of after-merger frequency of gravitational waves}\label{appfinalfeqn}

We can also attempt to estimate the frequency of gravitational waves soon after merger has taken place.
For this, we begin {\em assuming} angular momentum is nearly conserved around the moment where
the collision (contact) takes place. For the case of irrotational binaries, the angular momentum at contact can be approximated by the expression
\begin{equation}
    L_{c} = I_c \Omega_c ~~,
\end{equation}
with $I_c$ and $\Omega_c$ as defined in Appendix~\ref{appA}. Soon after contact, we have instead 
\begin{equation}
   L_{r} = I_r \Omega_r ~~,
\end{equation}
where $I_r$ is the moment of inertia of the newly formed object
(i.e, the remnant) assuming no prompt collapse to a BH
takes place. Let us now assume a relation of these angular momentum given by $L_r = \kappa L_c$ with $\kappa\in[0,1]$
a factor introduced to account for loss of angular momentum during the merger.
Now, since the angular frequency of gravitational waves $\omega  = 2\Omega$, we have (specializing to the
equal mass case)
\begin{eqnarray}
I_r \omega_r &=& \kappa 2 M R^2 \omega_c  ~~, \nonumber \\
M_r^3 (I_r/M_r^3) \omega_r &=& \kappa 2 M^3 C^{-2} \omega_c ~~,
\end{eqnarray}
with $C$ denoting the compactness of the individual stars, $M$ their individual masses, and $M_r$ and $I_r$ the 
mass of the remnant and its moment of inertia respectively soon after contact. Rearranging, the gravitational wave frequency 
soon after contact implies
\begin{equation}
\omega_r =  \frac{\kappa}{4} \frac{C^{-2}}{\bar I_r} \omega_c
\label{luis_mom}
\end{equation}
(where we have defined $\bar I_r = I_r M_r^{-3}$).

In the case of binary neutron star mergers, extensive studies already indicate a small amount of angular
momentum is radiated during the merger, so we can adopt $\kappa \simeq 1$. For realistic equations of
state, we can make use of thorough investigations of the values of $\bar I_r$ for isolated stars for a wide
range of compactness and mass (e.g.~\cite{Yagi:2013awa}) to
evaluate Eq.~(\ref{luis_mom}). This gives $\omega_r \approx 2.8 \, \omega_c$, 
in excellent agreement with results from numerical
simulations (e.g.~\cite{2016CQGra..33r4002L}). This is not surprising since the angular momentum in the system
right before contact is primarily transferred to the object formed after merger.

This is not the case for binary boson stars. As opposed to the case for neutron stars,
general values of $\bar I_r$ for boson stars are not available but we
can estimate them either from our isolated BS solutions or by considering that
they behave as constant density spheres, namely
$I=(2/5) M R^2$. Since the result will not change significantly, we use the latter approximation which yields
\begin{eqnarray}
  \omega_r = \kappa \frac{5}{2} \left(\frac{R}{R_f} \right)^2 \omega_c \approx 2\,\kappa\, \omega_c ~~,
\end{eqnarray}
where again we estimated $R/R_f \approx 0.9$ from our simulations. 
The corresponding estimate would give an upper bound factor (taking $\kappa=1$) in $[2-2.5]$ which
is higher than the fit in Eq.~(\ref{bsfit}). However, as we have seen, 
a large amount of angular momentum is lost through the merger. From Fig.~\ref{admmass} one would expect
$\kappa \approx 1/4$ making the ``expected'' frequencies from this naive estimate much too low when compared
with the measured peak frequency in gravitational waves. This is in strong contrast to what is observed in the case
of binary neutron star mergers.

As the next section illustrates, the after-merger radiation of boson stars 
is determined by the quasi-normal modes of the produced boson star or by the BH in the
case of collapse. This is a consequence of our observation that the merger of boson stars does not produce
a rotating boson star and that the speed of propagation of perturbations in boson stars is faster than that
in neutron stars.

\section{Quasi-normal modes of isolated solitonic boson star}\label{qnm_isolated}

\begin{figure}
\centering
\includegraphics[width=1.0\linewidth]{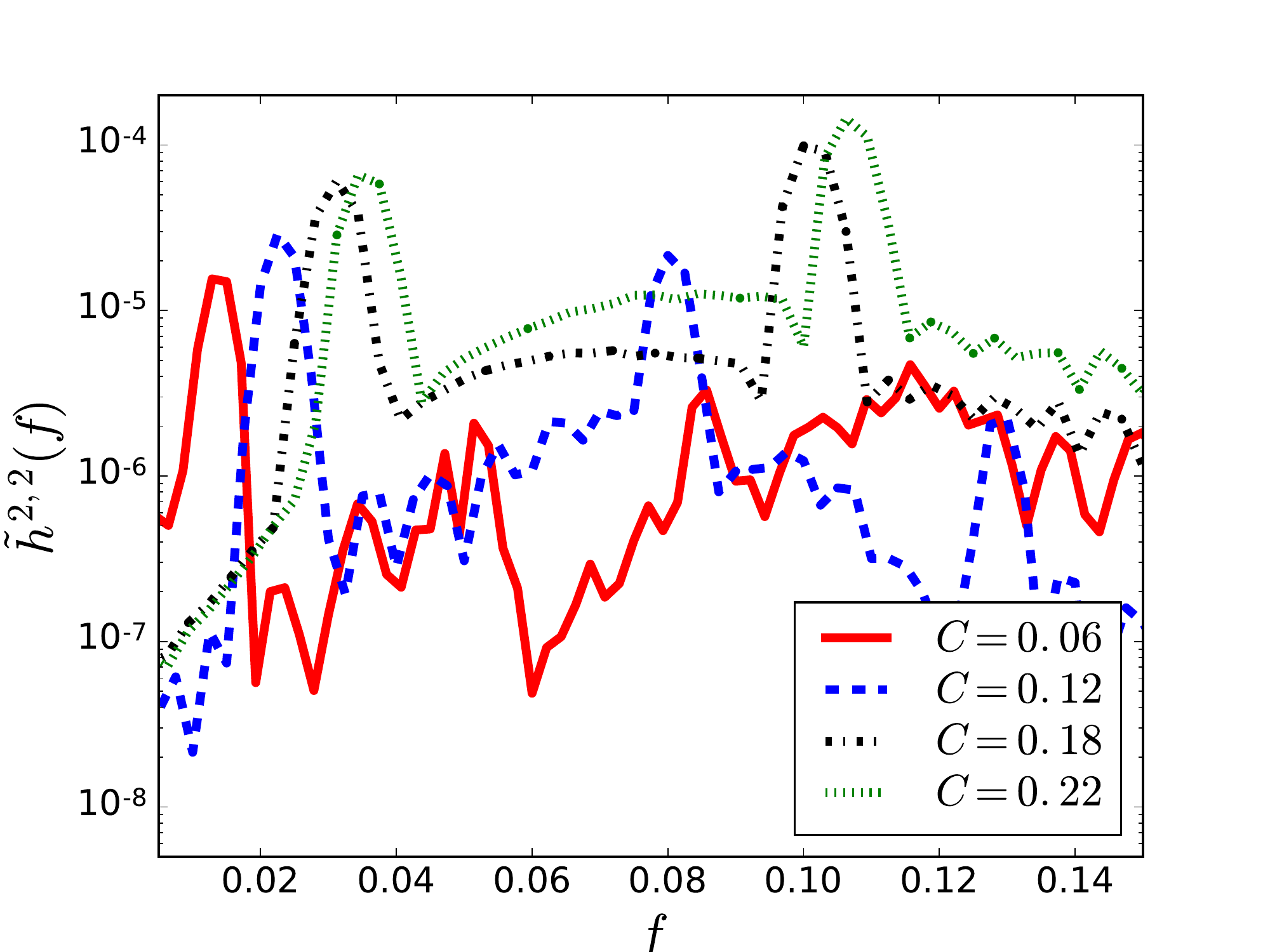}
\includegraphics[width=1.0\linewidth]{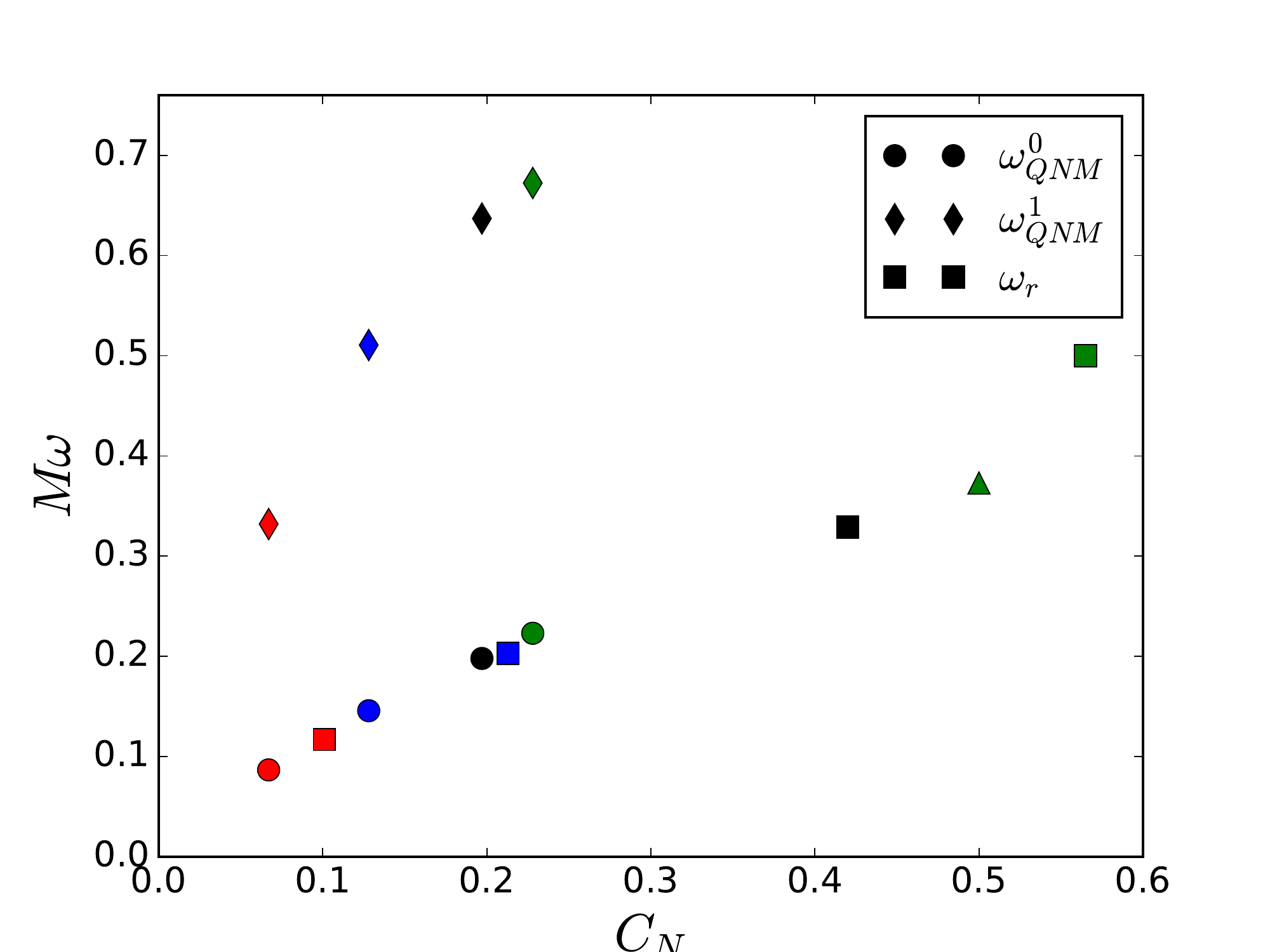}
\caption{(Top) Fourier spectrum of the main mode of the strain 
${\tilde h}^{2,2}(f) \equiv {\cal F} [h_{22}(t)]$ for several isolated solitonic boson stars. (Bottom) The circles and the diamonds correspond,respectively, to the frequencies $\omega^0_\mathrm{QNM}$ and $\omega^1_\mathrm{QNM}$ of the lowest quasi-normal modes (i.e., fundamental and secondary peaks), as a function of the compactness $C_N \equiv M/R_N$. Squares represent the gravitational frequencies of the remnant resulting from a binary merger with that initial compactness.
Notice that we have included the case with $C=0.18$, which we do not
trust completely, and the case $C=0.22$ that ends up in Kerr BH. For comparison purposes, we have included also the QNM of a Schwarzschild BH (triangle).
}
\label{qnms_iso}
\end{figure}

Quasi-normal modes for isolated solitonic boson stars can be computed
by evolving numerically a perturbed star and analyzing the gravitational wave radiation. 
The formalism and numerical schemes are the same as the ones used
in this work for binaries, such that only the initial data differs.
We have chosen stable boson stars with total mass $M=1$ for different compactnesses ranging from $C=0.06-0.22$. These equilibrium configurations are deformed by adding a small  perturbation on the conformal
factor which introduces {constraint violations} below the truncation error of the unperturbed configuration, and so we need not re-solve the constraints.
In order to ensure the excitation of gravitational modes,
the perturbation has a toroidal shape with a $m=2$ dependence in the axial direction. 

The top panel of Fig.~\ref{qnms_iso} shows the Fourier transform of the main gravitational wave mode (i.e., $l=m=2$) as a function of frequency. Although there are several peaks in the spectra, we focus only on the two strongest modes at the lowest frequencies. Clearly, the frequencies of the fundamental and the secondary quasi-normal modes increase with compactness. The bottom panel displays the adimensional frequency of these two modes as a function of the compactness. We have also included the frequencies of the remnant after the merger of the case studied here, given in Table \ref{table2}. There
is a  good agreement between the QNM of the single stars and the fundamental mode of the remnant of the binary.

\bibliographystyle{utphys}
\bibliography{biblio}

\end{document}